\documentclass[a4paper,onecolumn,11pt,unpublished]{quantumarticle}
\pdfoutput=1
 \usepackage[utf8]{inputenc}
\usepackage{amsmath,amssymb,amsthm,mathtools}
\usepackage{amsmath,amssymb}
\usepackage[hidelinks]{hyperref}
\usepackage{xcolor}
\usepackage{quantikz}
\usepackage{graphicx}
\usepackage{xcolor}
\usepackage{braket}
\usepackage[
  backend=biber,
  style=numeric,
  sorting=none      
]{biblatex}
\usepackage{algorithm}
\usepackage{algpseudocode}

\usepackage[T1]{fontenc}
\usepackage{setspace}              
\usepackage{bm}
\usepackage{amssymb}             
\usepackage{tikz}            
\usepackage{quantikz} 
\usepackage{enumitem}
\usepackage{graphicx} 
\usepackage{algpseudocode}
\usepackage{algorithm}
\usepackage{algorithmicx}
\usepackage{tikz}
\usetikzlibrary{positioning,arrows.meta,fit,calc}
\usepackage{float}

\usepackage{graphicx}

\usepackage[utf8]{inputenc}
\usepackage[english]{babel}
\usepackage{amssymb}
\usepackage{physics}

\usepackage{hyperref}
\usepackage[table]{xcolor}
\usepackage{colortbl}
\usepackage{booktabs}
\usepackage{tabularx}
\usepackage{caption}
\usepackage{subcaption}

\usepackage{selinput}
\SelectInputMappings{
  adieresis={ä}, 
  acute={´},
}

\usepackage[english]{babel}
\usepackage[T1]{fontenc}
\usepackage{lipsum}
\usepackage{mdframed}
\usepackage{booktabs}
\usepackage{adjustbox}
\usepackage{longtable}
 
\usepackage{placeins}
\newcolumntype{L}{>{\raggedright\arraybackslash}X}
\newcolumntype{C}{>{\centering\arraybackslash}X}
\newcolumntype{R}{>{\raggedleft\arraybackslash}X}

\usepackage{xcolor}
\usepackage[table]{xcolor}
\usepackage{booktabs}
\usepackage{listings}
\usepackage{siunitx}
\usepackage{multicol}
\usepackage{pifont}  
\usepackage{newunicodechar}
\newunicodechar{✖}{\ding{55}}
\usepackage{pdflscape}  
\usepackage{ragged2e}
\usepackage[most]{tcolorbox}
\usepackage{pgfplots}
\pgfplotsset{compat=1.17}

\usepackage{mathrsfs}
\newtheorem{proposition}{Proposition}

\usepackage{booktabs}
\usepackage{float}                 
\usepackage{hyperref}              
\usepackage{url}
\usepackage{xcolor}                
\usepackage{lscape}                
\usepackage{caption}
\usepackage{subcaption}
\usepackage{etoolbox}
\usepackage{rotating}
\usepackage{multirow}
\usepackage{xcolor}
\usepackage{adjustbox}
\pgfplotsset{compat=1.17}

\AtBeginEnvironment{abstract}{\raggedright}
\AtBeginDocument{\RenewCommandCopy\qty\SI} 
\DeclareUnicodeCharacter{0301}{'} 

\title{NEQSST: A Nested Quantum Sampling  Algorithm for the Shared Transportation Problem}

\author{Chinonso Onah}
\affiliation{Volkswagen AG, Berliner Ring 2, Wolfsburg 38440, Germany}
\affiliation{Department of Physics, RWTH Aachen University, Germany}
\email{Corr. Author. conah@atom-computing.com}

\author{Agneev Guin}
\affiliation{Volkswagen AG, Berliner Ring 2, Wolfsburg 38440, Germany}
\affiliation{
Department of Network Engineering, BarcelonaTech (UPC), Spain}
\email{agneev.guin@volkswagen.de}

\author{J.~A.~Montañez-Barrera}
\affiliation{Jülich Supercomputing Centre, Forschungszentrum Jülich, Germany}
\email{j.montanez-barrera@fz-juelich.de}

\author{Carsten Othmer}
\affiliation{Volkswagen AG, Berliner Ring 2, Wolfsburg 38440, Germany}
\email{carsten.othmer@volkswagen.de}

\author{Kristel Michielsen}
\affiliation{Department of Physics, RWTH Aachen University, Germany}
\affiliation{Jülich Supercomputing Centre, Forschungszentrum Jülich, Germany}
\email{k.michielsen@fz-juelich.de}

\begin{document}
\maketitle
\begin{abstract}
Shared transportation over a single road segment can be formulated as a
maximum-weight matching problem between drafting vehicles, referred to as
surfers, and leading vehicles, referred to as breakers. A multi-segment route introduces additional structure.  Vehicles may enter or leave at intermediate junctions, assignments must remain feasible on every segment, and transitions between adjacent segments create handover costs. We formulate this setting as a  segment-coupled quadratic binary optimization problem and introduce \textsc{NEQSST}, a nested quantum sampling algorithm built from one-hot segment--breaker blocks. The encoding enforces breaker capacity inside the encoded Hilbert space, removes the corresponding penalty family from the implemented phase Hamiltonian, and leaves surfer uniqueness and adjacent-segment
handover terms as commuting diagonal couplings between blocks. We derive the QUBO and Ising representations, give a blockwise decoder, and analyze the resulting qubit and phase-term counts. Exact nested optimization agrees with the direct Gurobi formulation on all tested instances, while finite-shot Qiskit simulation recovers the same optima for the reported problems. The results establish a consistent route from multi-segment shared-transportation modeling to an encoded quantum sampling implementation.
\end{abstract}

\section{Introduction}
\label{sec:introduction}

The shared transportation problem coordinates vehicles that travel along overlapping parts of a route to optimize for various objectives. In the Windbreaking-as-a-Service (WaaS) introduced in \cite{onah2025quest, onah2026solverzoo}, a leading vehicle, called a breaker, can reduce the effective energy cost of a following vehicle, called a surfer through aerodynamic drag reduction\cite{alam2015review,duoba2023empirical}. On a single road segment, the assignment problem has the structure of a weighted bipartite matching with side constraints where each surfer can select at most one breaker, and each breaker can serve at most one surfer\cite{onah2025quest, onah2026solverzoo}. The quality of an assignment is determined by the resulting energy saving together with departure-time and speed compatibility. 

A route containing several road segments introduces a coupled optimization problem. Vehicles may enter or leave at different junctions, so the set of admissible surfer--breaker pairs changes from one segment to the next. Assignments on adjacent segments are also linked by handover costs. A sequence of independently optimal segmentwise matchings can therefore have a smaller global objective than a schedule that sacrifices local reward to preserve a favorable continuation pattern. This paper develops a complete multi-segment formulation and an encoded quantum sampling method for this setting. 

The classical decision variable
\(x_{s,b,k}\) records whether surfer \(s\) drafts behind breaker \(b\) on
segment \(k\). The resulting objective combines segmentwise assignment rewards, surfer-uniqueness constraints, breaker-capacity constraints, timing compatibility, and handover costs on the segment-connection graph. The formulation admits an explicit QUBO representation and a diagonal Ising cost Hamiltonian. The quantum construction reorganizes the binary search space into nested one-hot blocks. For every admissible segment--breaker pair \((k,b)\), one block \(Y_{k,b}\) stores either the no-drafting label or the identity of the surfer
assigned to that breaker on that segment. This representation enforces breaker capacity by construction. The corresponding penalty family vanishes on the encoded subspace, while surfer uniqueness and handover consistency remain as cross-block phase couplings. Block-local XY mixers preserve each one-hot sector and allow all segment--breaker blocks to be initialized and mixed in parallel.

We call the resulting method \textsc{NEQSST}, for \emph{Nested Quantum
Sampling for Shared Transportation}. The implementation combines route
preprocessing, classical validation, nested quantum-state preparation,
commuting diagonal phase separation, blockwise decoding, and finite-shot
sampling. The main contributions are:
\begin{enumerate}[label=(\roman*)]
    \item a multi-segment shared-transportation model with route availability, timing compatibility, and adjacent-segment handover costs;
    \item an explicit QUBO and Ising formulation of the coupled problem;
    \item a nested segment--breaker one-hot encoding that enforces breaker
    capacity inside the encoded search space;
    \item a CE-QAOA realization with parallel block preparation, block-local XY mixing, and a commuting diagonal cost phase; and
    \item a classical decoder for measured nested samples; and
    \item a resource analysis focused on near-term hardware implementations.
\end{enumerate}

Weighted assignment and maximum-weight matching provide the classical basis for single-segment shared-transportation coordination \cite{Burkard2009}. Binary quadratic formulations extend such models to
settings with collision penalties, compatibility costs, and coupled assignment  decisions. General QUBO and Ising constructions for combinatorial optimization are reviewed in Refs.~\cite{Lucas2014,Katzgraber2015}. Quantum alternating-operator methods use problem-dependent state spaces and mixers to preserve selected constraints during quantum evolution. One-hot encodings combined with XY-type mixers are especially natural for assignment
variables because the single-excitation sector represents exactly one local choice\cite{Hadfield2019AOA}. The CE-QAOA framework used here provides block-local state preparation, constraint-preserving transport, and encoded phase separation
\cite{onahce,onahfinite}.

The present work also extends the shared-transportation setting introduced in the QUEST formulation \cite{onah2025quest}. The multi-segment problem adds route-dependent availability and handover coupling across adjacent road segments. \textsc{NEQSST} represents these decisions through nested segment--breaker blocks and separates the constraints enforced by the encoding from the couplings retained in the cost Hamiltonian. The remainder of the paper proceeds as follows. Section~\ref{sec:single-segment} derives the single-segment energy-saving model. Section~\ref{sec:multisegment-model} introduces the
multi-segment objective and route preprocessing.
Section~\ref{sec:qubo-ising} derives the QUBO and Ising representations.
Section~\ref{sec:nqsst} presents the nested quantum sampling construction and decoder. Section~\ref{sec:resource-analysis} analyzes circuit structure and resource counts. Sections~\ref{sec:implementation} and \ref{sec:results} describe the implementation and numerical results.
Section~\ref{sec:discussion} discusses scope and limitations, and
Section~\ref{sec:conclusion} concludes the paper.

\section{Maximum Energy--Saving Formulation}
\label{sec:single-segment}

Define the \emph{base} edge weights
\begin{equation}
  w^{\mathrm{base}}_{s,b}
  \;=\;
  c_s\,v_s^2 \;-\; c_s\,V_b^2\,\bigl[\,1 - f\!\bigl(C_b - c_s\bigr)\bigr],
  \label{eq:new_edge}
\end{equation}
for surfer \(s\) and breaker \(b\).
Let \(x_{s,b}\in\{0,1\}\) indicate whether \(s\) drafts behind \(b\).
If everyone rides alone, the baseline energy is \(F^{\mathrm{alone}}=\sum_s c_s v_s^2\).
Under a matching \(x\), the total savings relative to this baseline are
\begin{equation}
  \Delta F(x)
  \;=\;
  \sum_{s,b} x_{s,b}\,\bigl(c_s\,v_s^2 \;-\; c_s\,V_b^2\,\bigl[1 - f(C_b - c_s)\bigr]\bigr)
  \;=\;
  \sum_{s,b} x_{s,b}\,w^{\mathrm{base}}_{s,b}.
\end{equation}
Hence maximizing savings is equivalent to a maximum weight matching:
\begin{align}
  \max_{x\in\{0,1\}^{S\times B}}
  &\quad \sum_{s,b} x_{s,b}\,w^{\mathrm{base}}_{s,b} \\
  \text{s.t.}\quad
  &\sum_{b} x_{s,b} \;\le\; 1 \quad \forall s,
  \qquad
  \sum_{s} x_{s,b} \;\le\; 1 \quad \forall b.
\end{align}
See \cite{Burkard2009}.

\subsection{QUBO Formulation}

Let \(\mathcal S=\{1,\dots,S\}\), \(\mathcal B=\{1,\dots,B\}\).
Each potential edge \((s,b)\) has a binary decision \(x_{s,b}\in\{0,1\}\).
To encode time/speed preferences, define penalties
\[
\Delta^{T}_{s,b} \;=\;
\begin{cases}
|t_s-T_b|, & \text{if } |t_s-T_b| > \delta t_s/2,\\
0, & \text{otherwise,}
\end{cases}
\qquad
\Delta^{V}_{s,b} \text{ defined analogously with } \delta v_s.
\]

Here \(\delta t_s\) and \(\delta v_s\) are surfer-individual flexibility intervals
for departure time and preferred travel speed, respectively.
Let \(\lambda_1,\lambda_2\ge 0\) be weights and set the \emph{penalized weights}
\begin{equation}
  \tilde w_{s,b}
  \;=\;
  w^{\mathrm{base}}_{s,b}
  \;-\; \lambda_1\,\Delta^{T}_{s,b}
  \;-\; \lambda_2\,\Delta^{V}_{s,b}.
  \label{eq:wtilde}
\end{equation}

We seek a matching-or-none per row and column:
\[
\sum_{b} x_{s,b}\le 1 \quad \forall s,
\qquad
\sum_{s} x_{s,b}\le 1 \quad \forall b.
\]

Using quadratic pairwise collision penalties with parameter \(\lambda_3>0\), the unconstrained maximization becomes
\begin{equation}
\max_{x\in\{0,1\}^{S\times B}}
\Biggl[
\sum_{s,b} \tilde w_{s,b}\,x_{s,b}
\;-\;
\lambda_3 \sum_{s}\sum_{b<b'} x_{s,b}\,x_{s,b'}
\;-\;
\lambda_3 \sum_{b}\sum_{s<s'} x_{s,b}\,x_{s',b}
\Biggr],
\label{eq:qubo-max}
\end{equation}
with $\lambda_3>\max_{s,b}(\tilde w_{s,b})_+$, which is sufficient to suppress collisions. Setting \(n:=S(=B)\) and vectorizing in row-major order,
\[
x \;=\; \bigl(
x_{1,1},\,x_{1,2},\,\dots,\,x_{1,n},\,
x_{2,1},\,\dots,\,x_{n,n}
\bigr)^{\!\top},
\]
the equivalent minimization QUBO form
\[
\min_{x\in\{0,1\}^{n^2}} x^\top Q x
\]
is obtained by negating the objective coefficients. Since for symmetric \(Q\),
\[
x^\top Q x
=
\sum_i Q_{ii}x_i
+
2\sum_{i<j}Q_{ij}x_ix_j,
\]
the matrix entries are
\begin{align}
\label{eq:minqubo}
Q_{(s,b),(s,b)} &= -\,\tilde w_{s,b},\\
Q_{(s,b),(s,b')} &= \tfrac{\lambda_3}{2}\quad (b<b'),\\
Q_{(s,b),(s',b)} &= \tfrac{\lambda_3}{2}\quad (s<s'),\\
Q_{(s,b),(s',b')} &= 0 \quad (s\neq s',\ b\neq b').
\end{align}

Compactly,
\begin{equation}
Q \;=\; \operatorname{Diag}\!\bigl(-\,\mathrm{vec}(\tilde W)\bigr)
\;+\; \frac{\lambda_3}{2}\,\Bigl[\, I_n \otimes (J_n - I_n) \;+\; (J_n - I_n) \otimes I_n \,\Bigr],
\label{eq:Q-compact}
\end{equation}
where \(I_n\) is the \(n\times n\) identity, \(J_n\) the all-ones matrix, \(\otimes\) the Kronecker product, \(\tilde W=(\tilde w_{s,b})\), and
\[
\mathrm{vec}(\tilde W)=\bigl(\tilde w_{1,1},\tilde w_{1,2},\dots,\tilde w_{1,n},\tilde w_{2,1},\dots,\tilde w_{n,n}\bigr)^{\!\top}.
\]

\section{General Multi  Segment Problem}
\label{sec:multisegment-model}

\subsection{Notation}

We use the following notations recalled from \cite{onah2025quest}:
\begin{itemize}
    \item $S$: number of windsurfers, indexed by $s = 1,\dots,S$.
    \item $B$: number of windbreakers, indexed by $b = 1,\dots,B$.
    \item $K$: number of highway segments, indexed by $k = 1,\dots,K$.
    \item $c_s$ is the vehicle numerical vehicle-class parameter associated with surfer $s$,
    \item $C_b$ is the vehicle numerical vehicle-class parameter associated with breaker $b$,
    \item $d_{s,k}$: distance that \emph{surfer} $s$ must travel to reach segment $k$.
    \item $D_{b,k}$: distance that \emph{breaker} $b$ must travel to reach segment $k$.
    \item $v_s$: speed of surfer $s$.
    \item $V_b$: speed of breaker $b$.

\end{itemize}

We define a binary variable $x_{s,b,k} \in \{0,1\}$ for each surfer--breaker pair $(s,b)$ and segment $k$ as follows:
\[
x_{s,b,k} = 1
\quad \Longleftrightarrow 
\]
\[
\text{``surfer $s$ is assigned to breaker $b$ on segment $k$''.}
\]

Let
\begin{align}
  Cost
  &=
  \underbrace{\sum_{k}\sum_{s}\sum_{b}
      \tilde w_{s,b,k}\,x_{s,b,k}}_{\text{assignment}}
  \;-\;
  \underbrace{\lambda_h \sum_{s}\sum_{\substack{k\sim k'\\ k<k'}}\sum_{b,b'}
      \Delta T_{b,b',k,k'}\,
      x_{s,b,k}\,x_{s,b',k'}}_{\text{handover}} .
\end{align}

To capture these possibilities we define the coefficient of this quadratic term as:

\begin{equation}
\Delta T_{b,b',k,k'} 
\;=\; 1 + 
\bigl|\,
T_{b,k}
\;-\;
T_{b',k'}
\bigr|.
\end{equation}

Before constructing the QUBO, we perform route preprocessing and retain
only triples in the set
\begin{equation}
\mathcal I
=\{(s,b,k): s\text{ and }b\text{ both traverse segment }k\}.
\label{eq:retained_triples}
\end{equation}
If a vehicle does not traverse segment $k$, all associated variables are
fixed to zero and omitted. Hence every sum over assignment triples is
restricted to $\mathcal I$, and arrival times are evaluated only where defined.
Timing-infeasible triples may either remain in $\mathcal I$ with the penalty
$P_3$ specified below, or be removed as an additional hard-preprocessing
step. We use the penalty convention in the algebraic formulation and ensure
that the sample decoder discards any timing-infeasible assignment.

Indicator of infeasible timing:
\[
  \delta_{s,b,k}=
  \begin{cases}
    0, & |t_{s,k}-T_{b,k}|\le \Delta,\\
    1, & \text{otherwise.}
  \end{cases}
\]

We impose the constraints
\[
  \sum_b x_{s,b,k}\le 1
  \qquad \forall\, s,k,
\]
and
\[
  \sum_s x_{s,b,k}\le 1
  \qquad \forall\, b,k,
\]
using pairwise collision penalties.

\paragraph{(a) Surfer uniqueness}
\[
  P_1
  =
  \lambda_{3}\sum_{s,k}\sum_{b<b'}
  x_{s,b,k}\,x_{s,b',k}.
\]

\paragraph{(b) Breaker capacity}
\[
  P_2
  =
  \lambda_{4}\sum_{b,k}\sum_{s<s'}
  x_{s,b,k}\,x_{s',b,k}.
\]

\paragraph{(c) Timing feasibility}
\[
  P_3
  =
  \lambda_{5}\sum_{s,b,k}\delta_{s,b,k}\,x_{s,b,k}.
\]

Thus the unconstrained objective is

\begin{equation}
\label{eq:multisegment-waas-objective}
\begin{aligned}
H(\mathbf{x})
  &= Cost - P_1 - P_2 - P_3 \\[2pt]
  &= \sum_{s,b,k}
      \bigl(\tilde w_{s,b,k} - \lambda_{5}\delta_{s,b,k}\bigr)x_{s,b,k} \\[2pt]
  &\quad
    - \lambda_{3}\sum_{s,k}\sum_{b<b'}
      x_{s,b,k}\,x_{s,b',k} \\[2pt]
  &\quad
    - \lambda_{4}\sum_{b,k}\sum_{s<s'}
      x_{s,b,k}\,x_{s',b,k} \\[2pt]
  &\quad
    - \lambda_{h} \sum_{s}\sum_{\substack{k\sim k'\\ k<k'}}\sum_{b,b'}
      \Delta T_{b,b',k,k'}\,x_{s,b,k}\,x_{s,b',k'} .
\end{aligned}
\end{equation}

Here \(k\sim k'\) denotes that segments \(k\) and \(k'\) are adjacent in the
segment-connection graph \(G_{\mathrm{seg}}=(\mathcal K,E_{\mathrm{seg}})\), i.e.,
\[
k\sim k' \iff \{k,k'\}\in E_{\mathrm{seg}}.
\]
This notation is useful when the admissible handovers are not restricted to a
simple linear chain, but instead follow a tree or more general graph structure. For the special case of a linear route, \(k\sim k'\) reduces to \(|k-k'|=1\), and
together with \(k<k'\) this is equivalent to \(k'=k+1\).

\subsection{Multi-Segment QUBO}
\label{sec:qubo-ising}

To derive the QUBO matrix, flatten the variables into
\[
  \mathbf{x}=\bigl(x_i\bigr)_{i\in\mathcal I}^{\!\top}
  \in\{0,1\}^{|\mathcal I|},
\]
where a fixed ordering of $\mathcal I$ determines the matrix indices. In the
dense case $|\mathcal I|=SBK$. Define the symmetric maximization QUBO
matrix $Q_{\max}$ by
\[
  H(\mathbf{x})=\mathbf{x}^{\top}Q_{\max}\mathbf{x}.
\]
Since
\begin{equation}
\label{eq:qubo-obj}
  \mathbf{x}^{\top}Q_{\max}\mathbf{x}
  =\sum_i (Q_{\max})_{ii}x_i
  +2\sum_{i<j}(Q_{\max})_{ij}x_ix_j,
\end{equation}

the nonzero entries of $Q_{\max}$ are
\begin{align}
(Q_{\max})_{(s,b,k),(s,b,k)}
  &= \tilde w_{s,b,k}-\lambda_{5}\delta_{s,b,k},
  \\
(Q_{\max})_{(s,b,k),(s,b',k)}
  &= -\tfrac{\lambda_{3}}{2},\qquad b<b',
  \\
(Q_{\max})_{(s,b,k),(s',b,k)}
  &= -\tfrac{\lambda_{4}}{2},\qquad s<s',
  \\
(Q_{\max})_{(s,b,k),(s,b',k')}
  &= -\tfrac{1}{2}\lambda_{h}\Delta T_{b,b',k,k'},
  \qquad k\sim k',\ k<k'.
\end{align}
All entries not listed above vanish; entries related by transposition agree.

\paragraph{Sufficient penalty magnitudes.}
Let
\begin{equation}
a_{s,b,k}:=\tilde w_{s,b,k}-\lambda_5\delta_{s,b,k},
\qquad(s,b,k)\in\mathcal I.
\label{eq:waas-effective-reward}
\end{equation}
We assume $\lambda_h\geq0$ and $\Delta T_{b,b',k,k'}\geq0$ for every
handover term; the coefficient defined above is indeed nonnegative.
Writing $(u)_+=\max\{u,0\}$, a sufficient, instance-dependent choice is
\begin{equation}
\quad
\lambda_3,\lambda_4>
\max_{(s,b,k)\in\mathcal I}(a_{s,b,k})_+.
\quad
\label{eq:waas-sufficient-collision}
\end{equation}
For an empty retained set no assignment variable is needed. If timing
feasibility is enforced by $P_3$, it is sufficient to choose
\begin{equation}
\quad
\lambda_5>
\max_{\substack{(s,b,k)\in\mathcal I\\\delta_{s,b,k}=1}}
(\tilde w_{s,b,k})_+,
\quad
\label{eq:waas-sufficient-timing}
\end{equation}
with the right-hand side defined as zero when no such triples exist.

\begin{proposition}[Exactness of the multi-segment penalties]
\label{prop:waas-penalty-exactness}
Assume nonnegative handover coefficients, nonnegative $\lambda_h$,
and the strict collision bounds \eqref{eq:waas-sufficient-collision}.
Every maximizer of $H$ satisfies segmentwise surfer uniqueness and
breaker capacity. If timing-infeasible assignments are retained and
\eqref{eq:waas-sufficient-timing} also holds, every maximizer satisfies
timing feasibility. In particular, the global maximum equals the optimum
of the original hard-constrained objective.
\end{proposition}
\begin{proof}
Fix an active assignment $i=(s,b,k)$ with $x_i=1$ and delete it,
leaving all other variables unchanged. Let
\[
r_i=\sum_{b'\ne b}x_{s,b',k},\qquad
c_i=\sum_{s'\ne s}x_{s',b,k}.
\]
Let $h_i(x)$ be the sum of the nonnegative handover coefficients
$\Delta T$ associated with active quadratic terms incident on $i$,
counted once each. Direct subtraction gives
\begin{equation}
H(x-e_i)-H(x)
=-a_i+\lambda_3r_i+\lambda_4c_i+\lambda_hh_i(x).
\label{eq:waas-deletion-gain}
\end{equation}
If $i$ participates in either type of collision, then $r_i+c_i\geq1$,
so
\[
H(x-e_i)-H(x)\geq-a_i+\min\{\lambda_3,\lambda_4\}>0.
\]
Thus a violating assignment cannot occur in any maximizer.
If $i$ is timing-infeasible and \eqref{eq:waas-sufficient-timing}
holds, then $a_i<0$, and deletion strictly improves the objective even
without a collision. Repeated deletion proves feasibility; the removed
penalty terms vanish on the feasible set, where $H$ equals the
original hard-constrained objective.
\end{proof}

The local penalty bounds are needed in place of a bound based on the
absolute value of the global signed coefficient sum: cancellation in
that sum does not certify collision suppression.

\paragraph{Maximization versus minimization QUBO.}
The entries above represent the maximization objective $H$.
The conventional minimization matrix is
\begin{equation}
Q_{\min}:=-Q_{\max},\qquad
\operatorname*{arg\,max}_{x\in\{0,1\}^{|\mathcal I|}}
  x^\top Q_{\max}x
=
\operatorname*{arg\,min}_{x\in\{0,1\}^{|\mathcal I|}}
  x^\top Q_{\min}x.
\label{eq:waas-qubo-sign}
\end{equation}
The cost Hamiltonian below uses the maximization convention
$Q_{\max}$; changing its sign can equivalently be absorbed into the
phase-angle convention.

\section{Hybrid Quantum Classical Solution}
\label{sec:hybrid-qc}

\subsection{Multi-Segment Constraint-Enhanced QAOA}
\label{sec:ce-qaoa-multisegment}

The multi-segment QUBO can be implemented in the CE-QAOA framework by using a
nested one-hot encoding. For each road segment \(k\) and each breaker \(b\) that
traverses that segment, we introduce one block \(Y_{k,b}\). The states of this
block encode the possible surfer choices for breaker \(b\) on segment \(k\),
together with a dummy state \(0\) representing no drafting. Thus the CE-QAOA bitstring is naturally decomposed into segment-level groups,
breaker-level blocks, and surfer-choice states.

Let \(\mathcal K\) denote the set of road segments. After classical
preprocessing, let \(\mathcal B_k\) be the set of breakers that traverse segment
\(k\), and let \(\mathcal S_{k,b}\) be the set of surfers that can admissibly
draft behind breaker \(b\) on segment \(k\). For each admissible segment--breaker
pair \((k,b)\), we introduce one one-hot block
\[
Y_{k,b}
=
\{y_{k,b,0}\}\cup \{y_{k,b,s}:s\in\mathcal S_{k,b}\},
\]
where the label \(0\) denotes the no-drafting state. The one-hot constraint on
each block is
\[
\sum_{j\in\{0\}\cup\mathcal S_{k,b}} y_{k,b,j}=1.
\]
The original binary assignment variable is recovered from the block encoding by
\[
x_{s,b,k}=y_{k,b,s},
\qquad s\in\mathcal S_{k,b}.
\]
Thus each measured bitstring decomposes into segment-level groups, breaker-level
blocks, and surfer-choice states.

\begin{proposition}[Exact equivalence of the nested encoding]
\label{prop:waas-encoding-equivalence}
For each retained pair $(k,b)$ define the label set
$\mathcal Y_{k,b}=\{0\}\cup\mathcal S_{k,b}$ and
\[
\mathcal Y
=\prod_{k\in\mathcal K}\prod_{b\in\mathcal B_k}
\mathcal Y_{k,b}.
\]
Let
\[
\mathcal X_{\mathrm{cap}}
=\left\{x\in\{0,1\}^{|\mathcal I|}:
\sum_{s:(s,b,k)\in\mathcal I}x_{s,b,k}\leq1
\ \text{for all }(k,b)\right\}.
\]
Then $\mathscr D:\mathcal Y\to\mathcal X_{\mathrm{cap}}$ defined by
\begin{equation}
x_{s,b,k}=\mathbf1_{\{y_{k,b}=s\}}
\label{eq:waas-nested-bijection}
\end{equation}
is a bijection. The breaker-capacity penalty $P_2$ vanishes identically
on the nested encoded space. Under the assumptions of
Proposition~\ref{prop:waas-penalty-exactness},
\begin{equation}
\max_{y\in\mathcal Y}H_{\mathrm{enc}}(y)
=
\max_{x\in\mathcal F}\operatorname{Cost}(x),
\label{eq:waas-encoding-optima}
\end{equation}
where $\mathcal F$ is the set of fully feasible assignments, including
any hard timing requirements.
\end{proposition}
\begin{proof}
Each one-hot block has exactly one selected label. A dummy label
corresponds to no assignment, and a surfer label corresponds to precisely
one active variable for that breaker and segment. Conversely, every binary
assignment satisfying breaker capacity uniquely determines the selected
label in each block. These maps are inverses, proving the bijection.
Since at most one $x_{s,b,k}$ is active for fixed $(k,b)$,
\[
P_2(x)=\lambda_4\sum_{b,k}\sum_{s<s'}
 x_{s,b,k}x_{s',b,k}=0
\quad\text{on }\mathcal X_{\mathrm{cap}}.
\]
The encoded objective therefore agrees with the full penalized binary
objective at every decoded word. The penalty-exactness proposition shows
that a global maximizer is feasible with respect to the remaining
constraints. Every feasible assignment belongs to the encoded space and
has vanishing penalties, so the two optimal values coincide.
\end{proof}

In the dense case with \(K\) segments, \(B\) breakers per segment, and \(S\)
actual surfer choices per breaker block, each block contains the $S$ surfer
labels and one no-drafting label. The dense register therefore uses
\[
N_{\mathrm{qubits}}^{\mathrm{dense}}=KB(S+1).
\]
With preprocessing, the actual number of qubits is
\[
N_{\mathrm{qubits}}
=
\sum_{k\in\mathcal K}
\sum_{b\in\mathcal B_k}
\bigl(1+|\mathcal S_{k,b}|\bigr).
\]

Each block is initialized in the one-hot \(W\)-state
\[
\ket{s_{\mathrm{blk}}^{(k,b)}}
=
\frac{1}{\sqrt{1+|\mathcal S_{k,b}|}}
\sum_{j\in\{0\}\cup\mathcal S_{k,b}}
\ket{e_j},
\]
where \(\ket{e_j}\) is the computational-basis state with a single excitation
at label \(j\). The global initial state is the product state
\[
\ket{s_0}
=
\bigotimes_{k\in\mathcal K}
\bigotimes_{b\in\mathcal B_k}
\ket{s_{\mathrm{blk}}^{(k,b)}}.
\]

On each block we use a complete-graph XY mixer. With
$d_{k,b}=1+|\mathcal S_{k,b}|$, define
\begin{equation}
\widetilde H_{XY}^{(k,b)}
=
\begin{cases}
\displaystyle
\frac{1}{d_{k,b}-1}
\sum_{a<a'}(X_aX_{a'}+Y_aY_{a'}),&d_{k,b}\geq2,\\[6pt]
0,&d_{k,b}=1.
\end{cases}
\label{eq:waas-mixer-singleton}
\end{equation}
A singleton block contains only the no-drafting label and may equivalently
be omitted during preprocessing.
The CE-QAOA mixer is the product of block-local constrained
evolutions
\[
U_M(\beta)
=
\bigotimes_{k\in\mathcal K}
\bigotimes_{b\in\mathcal B_k}
\exp\!\left[-i\beta\,\widetilde H_{XY}^{(k,b)}\right].
\]
This mixer preserves the one-hot subspace in every \((k,b)\)-block. Therefore,
for each breaker \(b\) on each segment \(k\), the encoded state always represents
exactly one local choice: either no drafting, or one surfer assigned to that
breaker on that segment. The diagonal cost unitary is
\[
U_C(\gamma)
=
\exp[-i\gamma H_C],
\]
where \(H_C\) is obtained by promoting the QUBO matrix entries to diagonal number operators.

\paragraph{Cost Hamiltonian from the QUBO objective in Eq. \eqref{eq:qubo-obj}.}
\label{par:cost-hamiltonian-from-qubo}

Let \(i=(s,b,k)\) denote a flattened surfer--breaker--segment index, and let
\[
\widehat x_i
=
\widehat x_{s,b,k}
=
\frac{I-Z_{s,b,k}}{2}
\]
be the computational-basis number operator associated with the binary decision
variable \(x_{s,b,k}\). The QUBO objective
\[
H(\mathbf x)=\mathbf x^\top Q_{\max}\mathbf x
=
\sum_i (Q_{\max})_{ii}x_i+2\sum_{i<j}(Q_{\max})_{ij}x_ix_j
\]
is promoted to the diagonal quantum cost Hamiltonian
\begin{equation}
\label{eq:waas-ising-cost-general}
H_C
=
\sum_i (Q_{\max})_{ii}\widehat x_i
+
2\sum_{i<j}(Q_{\max})_{ij}\widehat x_i\widehat x_j .
\end{equation}
Substituting the multi-segment WaaS QUBO coefficients gives
\begin{equation}
\label{eq:waas-ising-cost}
\begin{aligned}
H_C
&=
\sum_{s,b,k}
\bigl(\tilde w_{s,b,k}-\lambda_5\delta_{s,b,k}\bigr)
\widehat x_{s,b,k}
\\[2pt]
&\quad
-
\lambda_3
\sum_{s,k}\sum_{b<b'}
\widehat x_{s,b,k}\widehat x_{s,b',k}
\\[2pt]
&\quad
-
\lambda_4
\sum_{b,k}\sum_{s<s'}
\widehat x_{s,b,k}\widehat x_{s',b,k}
\\[2pt]
&\quad
-
\lambda_h
\sum_s
\sum_{\substack{k\sim k'\\ k<k'}}
\sum_{b,b'}
\Delta T_{b,b',k,k'}
\widehat x_{s,b,k}\widehat x_{s,b',k'} .
\end{aligned}
\end{equation}
The CE-QAOA cost unitary is then
\begin{equation}
\label{eq:waas-cost-unitary}
U_C(\gamma)
=
\exp[-i\gamma H_C].
\end{equation}
Equivalently, Eq.~\eqref{eq:waas-ising-cost-general} may be expanded in
Pauli-\(Z\) form as
\begin{equation}
\label{eq:waas-pauli-z-cost}
H_C
=
c_0 I
+
\sum_i h_i Z_i
+
\sum_{i<j}J_{ij}Z_iZ_j,
\end{equation}
with
\begin{align*}
c_0&=\frac12\sum_i (Q_{\max})_{ii}
+\frac12\sum_{i<j}(Q_{\max})_{ij},\\
h_i&=-\frac12(Q_{\max})_{ii}
-\frac12\sum_{j\ne i}(Q_{\max})_{ij},\\
J_{ij}&=\frac12(Q_{\max})_{ij}.
\end{align*}
The constant shift \(c_0I\) contributes only a global phase to
\(U_C(\gamma)\). The implemented phase separator is therefore determined by the
one-body fields \(h_i\) and two-body couplings \(J_{ij}\).

For the nested CE-QAOA encoding, each segment--breaker block is one-hot and
therefore enforces breaker capacity inside the encoded Hilbert space. The
implemented encoded cost Hamiltonian is obtained by restricting
Eq.~\eqref{eq:waas-ising-cost} to the one-hot subspace of the CE-QAOA register;
on this subspace, the breaker-capacity penalty family proportional to
\(\lambda_4\) is identically zero and may be omitted from the phase
implementation \cite{onahce}.

A depth-\(p\) CE-QAOA state is therefore
\[
\ket{\psi_p(\boldsymbol\gamma,\boldsymbol\beta)}
=
\prod_{\ell=1}^{p}
U_M(\beta_\ell)U_C(\gamma_\ell)
\ket{s_0}.
\]
After measurement in the computational basis, a classical decoder partitions
the bitstring into its \((k,b)\)-blocks. If the active label in block
\(Y_{k,b}\) is \(0\), no surfer is assigned to breaker \(b\) on segment \(k\).
If the active label is \(s\), the decoder records the assignment
\((s,b,k)\). The resulting classical schedule is then evaluated using the
multi-segment objective, including the surfer uniqueness penalties and the
handover terms. The best decoded feasible schedule over all measured samples is returned as the WaaS solution


\definecolor{neqsstSOne}{HTML}{2F6BFF}
\definecolor{neqsstSTwo}{HTML}{E67E22}
\definecolor{neqsstSThree}{HTML}{2E9D56}
\definecolor{neqsstBreaker}{HTML}{7D3C98}
\definecolor{neqsstHandover}{HTML}{C0392B}

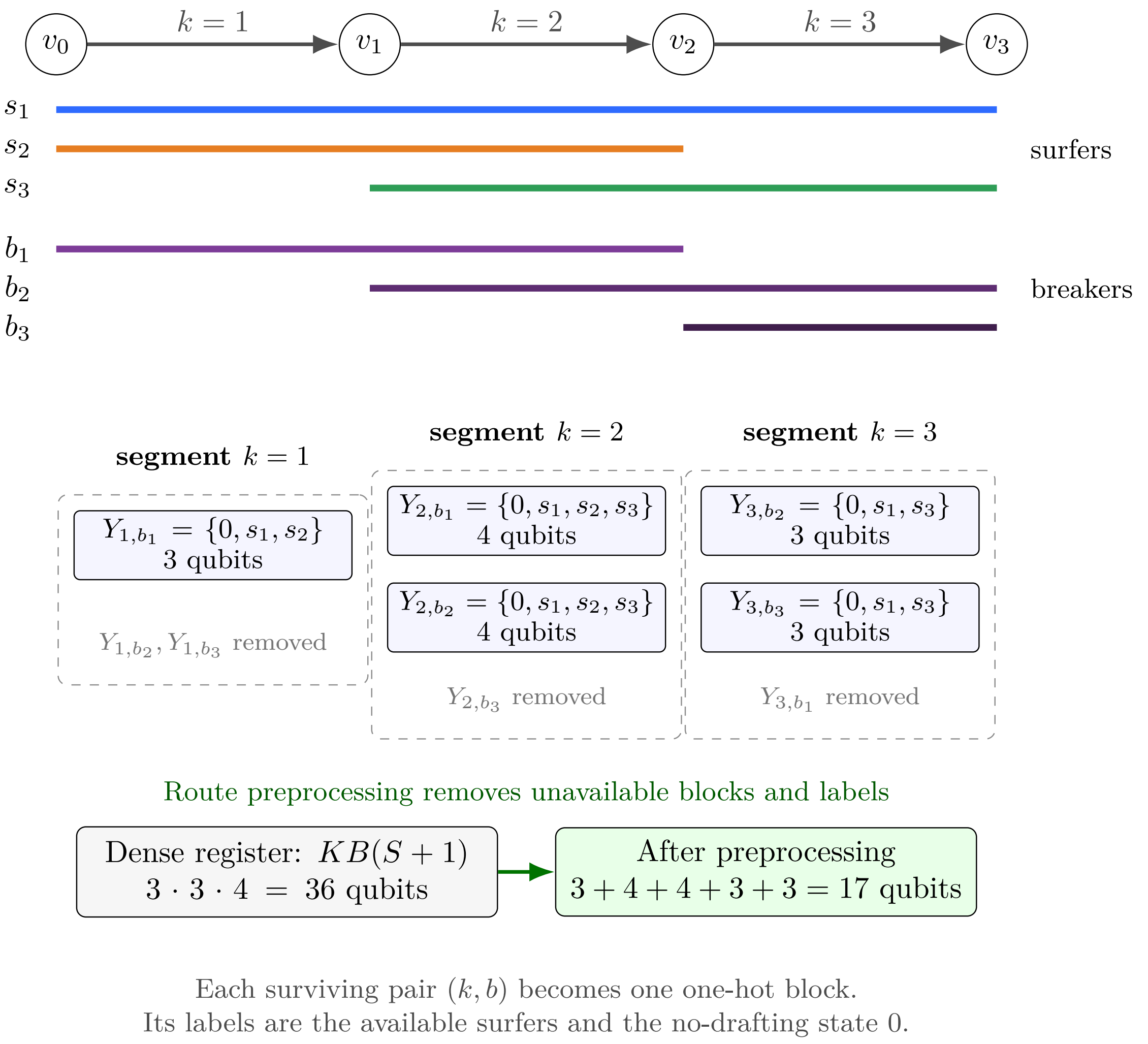
\begin{figure}[H]
\centering
\resizebox{\textwidth}{!}{%
\begin{tikzpicture}[
    >=Latex,
    junction/.style={circle,draw,fill=white,minimum size=7mm,inner sep=0pt},
    road/.style={->,very thick,black!70},
    route/.style={line width=2.2pt},
    regblock/.style={draw,rounded corners=2pt,fill=blue!4,
       text width=3.05cm,align=center,minimum height=0.79cm,inner sep=2pt,
       font=\footnotesize},
    seggroup/.style={draw,rounded corners=3pt,dashed,black!45,inner sep=5pt},
    summary/.style={draw,rounded corners=3pt,align=center,inner sep=4pt,
       text width=4.55cm,minimum height=0.9cm},
    every node/.style={font=\small}
]
\node[junction] (v0) at (0,6.50) {$v_0$};
\node[junction] (v1) at (3.6,6.50) {$v_1$};
\node[junction] (v2) at (7.2,6.50) {$v_2$};
\node[junction] (v3) at (10.8,6.50) {$v_3$};
\draw[road] (v0) -- node[above] {$k=1$} (v1);
\draw[road] (v1) -- node[above] {$k=2$} (v2);
\draw[road] (v2) -- node[above] {$k=3$} (v3);
\node[anchor=east] at (-0.15,5.75) {$s_1$};
\draw[route,neqsstSOne] (0,5.75) -- (10.8,5.75);
\node[anchor=east] at (-0.15,5.30) {$s_2$};
\draw[route,neqsstSTwo] (0,5.30) -- (7.2,5.30);
\node[anchor=east] at (-0.15,4.85) {$s_3$};
\draw[route,neqsstSThree] (3.6,4.85) -- (10.8,4.85);
\node[anchor=west,font=\footnotesize] at (11.05,5.30) {surfers};
\node[anchor=east] at (-0.15,4.15) {$b_1$};
\draw[route,neqsstBreaker] (0,4.15) -- (7.2,4.15);
\node[anchor=east] at (-0.15,3.70) {$b_2$};
\draw[route,neqsstBreaker!75!black] (3.6,3.70) -- (10.8,3.70);
\node[anchor=east] at (-0.15,3.25) {$b_3$};
\draw[route,neqsstBreaker!50!black] (7.2,3.25) -- (10.8,3.25);
\node[anchor=west,font=\footnotesize] at (11.05,3.70) {breakers};

\node[regblock] (y11) at (1.8,0.75)
  {$Y_{1,b_1}=\{0,s_1,s_2\}$\\[-1pt]$3$ qubits};
\node[font=\scriptsize,black!55] (r1) at (1.8,-0.40)
  {$Y_{1,b_2},Y_{1,b_3}$ removed};
\node[seggroup,fit=(y11)(r1)] (g1) {};
\node[above=4pt of g1,font=\footnotesize\bfseries] {segment $k=1$};

\node[regblock] (y21) at (5.4,1.03)
  {$Y_{2,b_1}=\{0,s_1,s_2,s_3\}$\\[-1pt]$4$ qubits};
\node[regblock] (y22) at (5.4,-0.08)
  {$Y_{2,b_2}=\{0,s_1,s_2,s_3\}$\\[-1pt]$4$ qubits};
\node[font=\scriptsize,black!55] (r2) at (5.4,-1.02)
  {$Y_{2,b_3}$ removed};
\node[seggroup,fit=(y21)(y22)(r2)] (g2) {};
\node[above=4pt of g2,font=\footnotesize\bfseries] {segment $k=2$};

\node[regblock] (y32) at (9.0,1.03)
  {$Y_{3,b_2}=\{0,s_1,s_3\}$\\[-1pt]$3$ qubits};
\node[regblock] (y33) at (9.0,-0.08)
  {$Y_{3,b_3}=\{0,s_1,s_3\}$\\[-1pt]$3$ qubits};
\node[font=\scriptsize,black!55] (r3) at (9.0,-1.02)
  {$Y_{3,b_1}$ removed};
\node[seggroup,fit=(y32)(y33)(r3)] (g3) {};
\node[above=4pt of g3,font=\footnotesize\bfseries] {segment $k=3$};


\node[font=\footnotesize,green!35!black] at (5.4,-2.10)
  {Route preprocessing removes unavailable blocks and labels};
\node[summary,fill=gray!7] (dense) at (2.65,-3.00)
  {Dense register: $KB(S+1)$\\$3\cdot3\cdot4=36$ qubits};
\node[summary,fill=green!9] (reduced) at (8.15,-3.00)
  {After preprocessing\\$3+4+4+3+3=17$ qubits};
\draw[->,very thick,green!45!black] (dense.east) -- (reduced.west);
\node[font=\footnotesize,align=center,black!70,text width=11cm]
  at (5.4,-4.55)
  {Each surviving pair $(k,b)$ becomes one one-hot block.\\
  Its labels are the available surfers and the no-drafting state $0$.};
\end{tikzpicture}%
}
\caption{Route-to-register construction for a three-segment instance with staggered vehicle entry and exit. Route availability determines which segment--breaker blocks survive preprocessing and which surfer labels appear in each one-hot block.}
\label{fig:route-to-register}
\end{figure}

\subsection{\textsc{NEQSST}: Nested Quantum Sampling for the Shared Transportation Problem}
\label{sec:nqsst}

We call the resulting algorithmic core \textsc{NEQSST}, for
\emph{Nested Quantum Sampling for Shared Transportation}. The method uses a nested one-hot representation of segment--breaker--surfer decisions, block-local constraint-preserving mixers, and a commuting diagonal cost phase to sample candidate shared-transportation schedules.

\begin{algorithm}[H]
\caption{Classical decoding of nested CE-QAOA WaaS samples}
\label{alg:ceqaoa-waas-decoder}
\begin{algorithmic}[1]
\Require 
A set of measured CE-QAOA bitstrings \(\mathcal Z\); segment set \(\mathcal K\);
preprocessed breaker sets \(\mathcal B_k\); admissible surfer sets
\(\mathcal S_{k,b}\); one-hot block map \(Y_{k,b}\); objective function \(H\).
\Ensure 
Best decoded WaaS schedule \(\mathcal M^\star\) and objective value
\(H^\star\).

\State \(H^\star \gets 0\) \Comment{The empty schedule is feasible}
\State \(\mathcal M^\star \gets \emptyset\)

\ForAll{measured bitstrings \(z\in\mathcal Z\)}
    \State \(\mathcal M \gets \emptyset\)
    \State \(x_{s,b,k}\gets 0\) for all admissible triples \((s,b,k)\)

    \ForAll{segments \(k\in\mathcal K\)}
        \ForAll{breakers \(b\in\mathcal B_k\)}
            \State Read the one-hot block \(Y_{k,b}\) from \(z\)
            \State Let \(j^\star\) be the active label in \(Y_{k,b}\)

            \If{\(j^\star = 0\)}
                \State Continue \Comment{No surfer drafts behind \(b\) on segment \(k\)}
            \Else
                \State \(s \gets j^\star\)
                \If{\(\delta_{s,b,k}=1\)}
                    \State \textbf{continue} \Comment{Discard hard timing violation}
                \EndIf
                \If{\(a_{s,b,k}\leq0\)}
                    \State \textbf{continue} \Comment{Never improves a feasible schedule}
                \EndIf
                \State \(x_{s,b,k}\gets 1\)
                \State \(\mathcal M \gets \mathcal M\cup\{(s,b,k)\}\)
            \EndIf
        \EndFor
    \EndFor

    \ForAll{segments \(k\in\mathcal K\)}
        \ForAll{surfers \(s\)}
            \State \(\mathcal C_{s,k}\gets \{b:(s,b,k)\in\mathcal M\}\)

            \If{\(|\mathcal C_{s,k}|>1\)}
                \State Select $b^\star\in\arg\max_{b\in\mathcal C_{s,k}}a_{s,b,k}$
                \State Remove all assignments \((s,b,k)\) with
                \(b\in \mathcal C_{s,k}\setminus\{b^\star\}\)
                \State Set \(x_{s,b,k}\gets 0\) for all removed assignments
            \EndIf
        \EndFor
    \EndFor

    \State Evaluate $H(z)\gets\operatorname{Cost}(x)$ using
           Eq.~\eqref{eq:multisegment-waas-objective} (all penalties vanish).

    \If{\(H(z)>H^\star\)}
        \State \(H^\star\gets H(z)\)
        \State \(\mathcal M^\star\gets \mathcal M\)
    \EndIf
\EndFor

\State \Return \((\mathcal M^\star,H^\star)\)
\end{algorithmic}
\end{algorithm}

\begin{figure}[htbp]
  \centering
  \resizebox{\textwidth}{!}{%
  \begin{quantikz}[row sep=0.1cm, column sep=0.05cm]
\lstick[wires=4]{$\text{Block }1\;(k=1,b=1)\; \ket{0}^{\otimes(S+1)}$} & \gate[wires=4,style={rounded corners,fill=blue!8}]{\texttt{OneHotBlock}} & \qw & \gate[wires=24,style={rounded corners,fill=green!16}]{U_C(\gamma_1)} & \gate[wires=4,style={rounded corners,fill=orange!20}]{U_M^{(0)}(\beta_1)} & \qw & \gate[wires=24,style={rounded corners,fill=green!16}]{U_C(\gamma_2)} & \gate[wires=4,style={rounded corners,fill=orange!20}]{U_M^{(0)}(\beta_2)} & \qw & \gate[wires=24,style={rounded corners,fill=green!16}]{U_C(\gamma_3)} & \gate[wires=4,style={rounded corners,fill=orange!20}]{U_M^{(0)}(\beta_3)} & \qw & \meter{} \\
 &  & \qw & \qw & \qw & \qw & \qw & \qw & \qw & \qw & \qw & \qw & \meter{} \\
 &  & \qw & \qw & \qw & \qw & \qw & \qw & \qw & \qw & \qw & \qw & \meter{} \\
 &  & \qw & \qw & \qw & \qw & \qw & \qw & \qw & \qw & \qw & \qw & \meter{} \\
\lstick[wires=4]{$\text{Block }2\;(k=1,b=2)\; \ket{0}^{\otimes(S+1)}$} & \gate[wires=4,style={rounded corners,fill=blue!8}]{\texttt{OneHotBlock}} & \qw & \qw & \gate[wires=4,style={rounded corners,fill=orange!20}]{U_M^{(1)}(\beta_1)} & \qw & \qw & \gate[wires=4,style={rounded corners,fill=orange!20}]{U_M^{(1)}(\beta_2)} & \qw & \qw & \gate[wires=4,style={rounded corners,fill=orange!20}]{U_M^{(1)}(\beta_3)} & \qw & \meter{} \\
 &  & \qw & \qw & \qw & \qw & \qw & \qw & \qw & \qw & \qw & \qw & \meter{} \\
 &  & \qw & \qw & \qw & \qw & \qw & \qw & \qw & \qw & \qw & \qw & \meter{} \\
 &  & \qw & \qw & \qw & \qw & \qw & \qw & \qw & \qw & \qw & \qw & \meter{} \\
\lstick[wires=4]{$\text{Block }3\;(k=1,b=3)\; \ket{0}^{\otimes(S+1)}$} & \gate[wires=4,style={rounded corners,fill=blue!8}]{\texttt{OneHotBlock}} & \qw & \qw & \gate[wires=4,style={rounded corners,fill=orange!20}]{U_M^{(2)}(\beta_1)} & \qw & \qw & \gate[wires=4,style={rounded corners,fill=orange!20}]{U_M^{(2)}(\beta_2)} & \qw & \qw & \gate[wires=4,style={rounded corners,fill=orange!20}]{U_M^{(2)}(\beta_3)} & \qw & \meter{} \\
 &  & \qw & \qw & \qw & \qw & \qw & \qw & \qw & \qw & \qw & \qw & \meter{} \\
 &  & \qw & \qw & \qw & \qw & \qw & \qw & \qw & \qw & \qw & \qw & \meter{} \\
 &  & \qw & \qw & \qw & \qw & \qw & \qw & \qw & \qw & \qw & \qw & \meter{} \\
\lstick[wires=4]{$\text{Block }4\;(k=2,b=1)\; \ket{0}^{\otimes(S+1)}$} & \gate[wires=4,style={rounded corners,fill=blue!8}]{\texttt{OneHotBlock}} & \qw & \qw & \gate[wires=4,style={rounded corners,fill=orange!20}]{U_M^{(3)}(\beta_1)} & \qw & \qw & \gate[wires=4,style={rounded corners,fill=orange!20}]{U_M^{(3)}(\beta_2)} & \qw & \qw & \gate[wires=4,style={rounded corners,fill=orange!20}]{U_M^{(3)}(\beta_3)} & \qw & \meter{} \\
 &  & \qw & \qw & \qw & \qw & \qw & \qw & \qw & \qw & \qw & \qw & \meter{} \\
 &  & \qw & \qw & \qw & \qw & \qw & \qw & \qw & \qw & \qw & \qw & \meter{} \\
 &  & \qw & \qw & \qw & \qw & \qw & \qw & \qw & \qw & \qw & \qw & \meter{} \\
\lstick[wires=4]{$\text{Block }5\;(k=2,b=2)\; \ket{0}^{\otimes(S+1)}$} & \gate[wires=4,style={rounded corners,fill=blue!8}]{\texttt{OneHotBlock}} & \qw & \qw & \gate[wires=4,style={rounded corners,fill=orange!20}]{U_M^{(4)}(\beta_1)} & \qw & \qw & \gate[wires=4,style={rounded corners,fill=orange!20}]{U_M^{(4)}(\beta_2)} & \qw & \qw & \gate[wires=4,style={rounded corners,fill=orange!20}]{U_M^{(4)}(\beta_3)} & \qw & \meter{} \\
 &  & \qw & \qw & \qw & \qw & \qw & \qw & \qw & \qw & \qw & \qw & \meter{} \\
 &  & \qw & \qw & \qw & \qw & \qw & \qw & \qw & \qw & \qw & \qw & \meter{} \\
 &  & \qw & \qw & \qw & \qw & \qw & \qw & \qw & \qw & \qw & \qw & \meter{} \\
\lstick[wires=4]{$\text{Block }6\;(k=2,b=3)\; \ket{0}^{\otimes(S+1)}$} & \gate[wires=4,style={rounded corners,fill=blue!8}]{\texttt{OneHotBlock}} & \qw & \qw & \gate[wires=4,style={rounded corners,fill=orange!20}]{U_M^{(5)}(\beta_1)} & \qw & \qw & \gate[wires=4,style={rounded corners,fill=orange!20}]{U_M^{(5)}(\beta_2)} & \qw & \qw & \gate[wires=4,style={rounded corners,fill=orange!20}]{U_M^{(5)}(\beta_3)} & \qw & \meter{} \\
 &  & \qw & \qw & \qw & \qw & \qw & \qw & \qw & \qw & \qw & \qw & \meter{} \\
 &  & \qw & \qw & \qw & \qw & \qw & \qw & \qw & \qw & \qw & \qw & \meter{} \\
 &  & \qw & \qw & \qw & \qw & \qw & \qw & \qw & \qw & \qw & \qw & \meter{} \\
\end{quantikz}}
  \caption{Nested CE-QAOA circuit for a two-segment WaaS instance with three breakers and three surfers. The six blocks correspond to segment--breaker pairs, each with three surfer labels and one dummy label (four qubits per block). The green cost unitary acts across the complete nested register, while the orange mixers act independently within the one-hot blocks.}
  \label{fig:two-segment-three-breaker-nested-ceqaoa}
\end{figure}
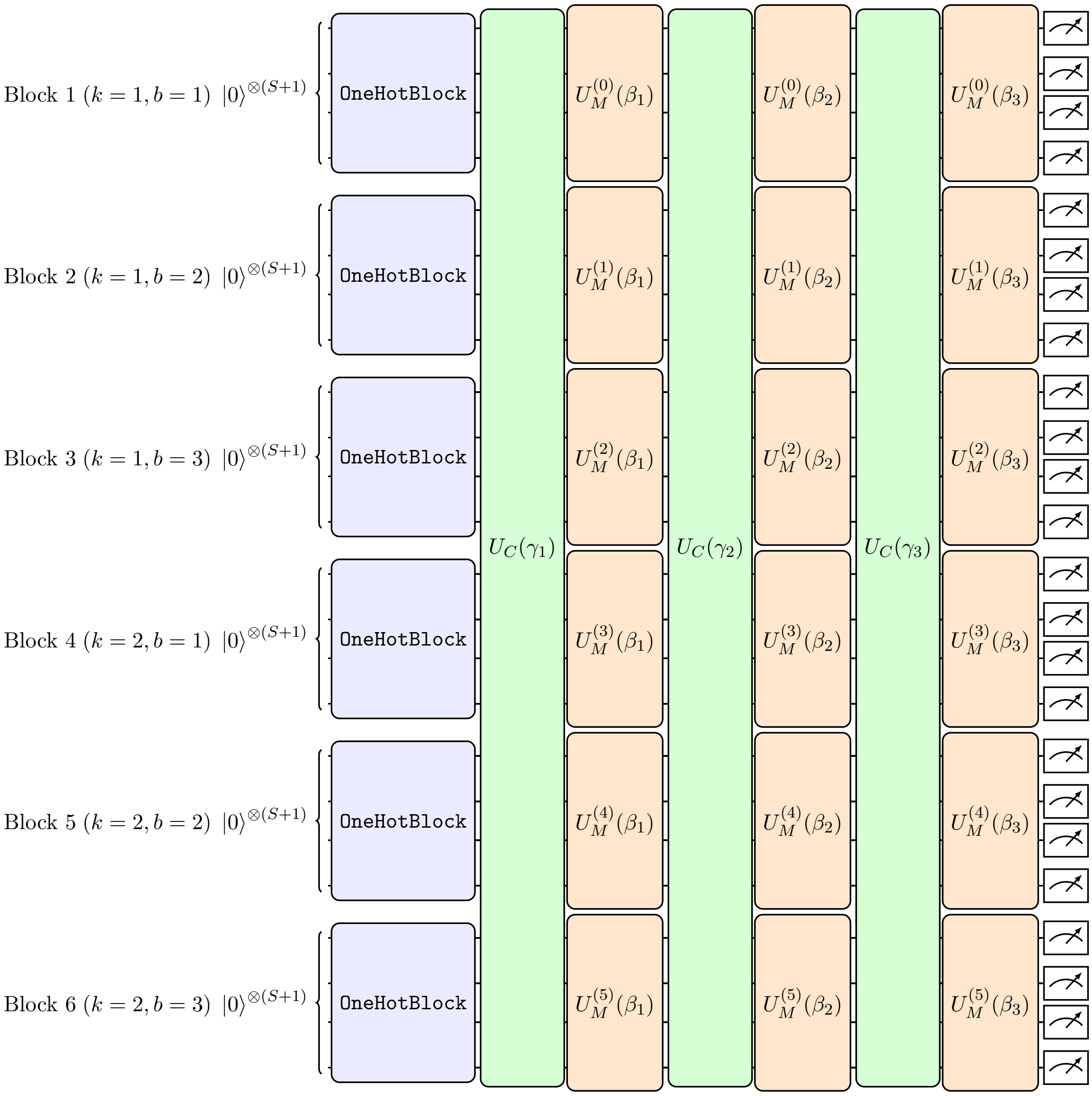

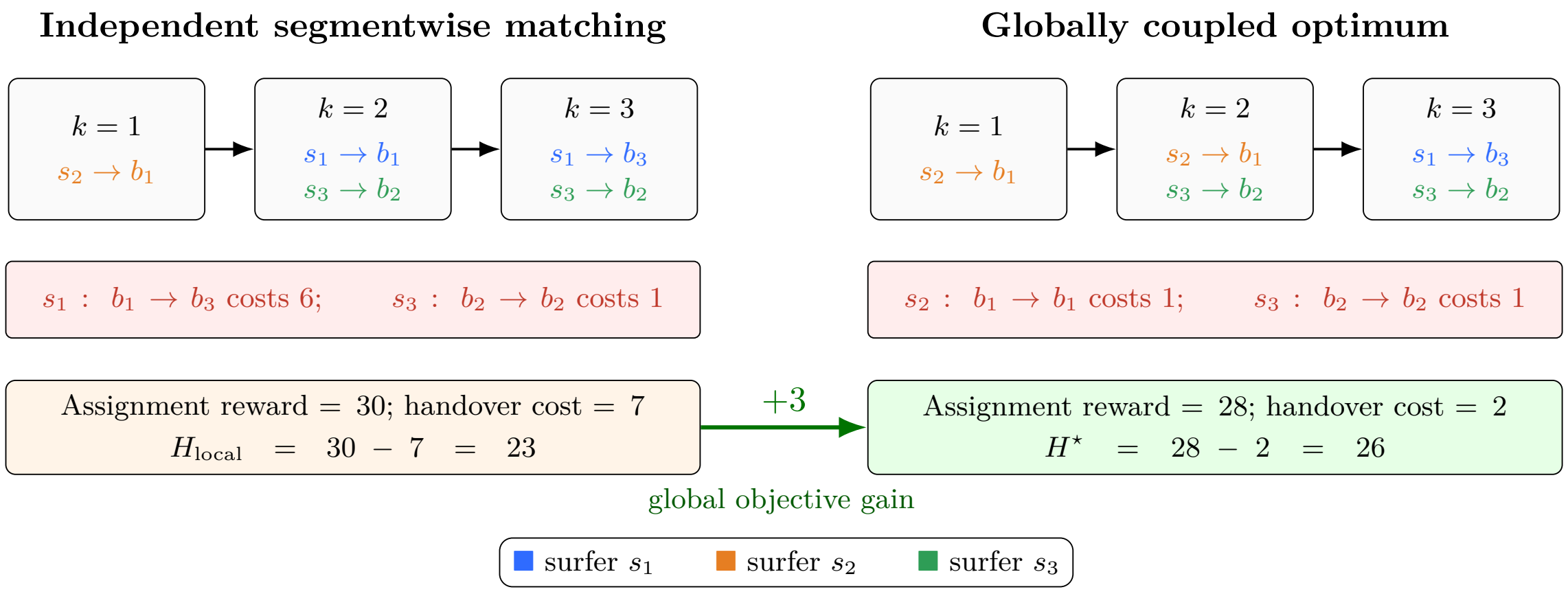
\begin{figure}[H]
\centering
\resizebox{\textwidth}{!}{%
\begin{tikzpicture}[
  >=Latex,
  segbox/.style={draw,rounded corners=3pt,minimum width=2.15cm,
     minimum height=1.55cm,align=center,fill=gray!4,font=\footnotesize},
  hand/.style={draw,rounded corners=2pt,fill=red!7,text=neqsstHandover,
     align=center,font=\footnotesize,text width=7.4cm,minimum height=.84cm,
     inner sep=3pt},
  score/.style={draw,rounded corners=3pt,align=center,text width=7.4cm,
     minimum height=1.03cm,inner sep=3pt,font=\footnotesize},
  every node/.style={font=\small}
]
\node[font=\bfseries] at (3.90,3.83) {Independent segmentwise matching};
\node[segbox] (l1) at (1.20,2.50)
 {$k=1$\\[3pt]\textcolor{neqsstSTwo}{$s_2\to b_1$}};
\node[segbox] (l2) at (3.90,2.50)
 {$k=2$\\[3pt]\textcolor{neqsstSOne}{$s_1\to b_1$}\\
  \textcolor{neqsstSThree}{$s_3\to b_2$}};
\node[segbox] (l3) at (6.60,2.50)
 {$k=3$\\[3pt]\textcolor{neqsstSOne}{$s_1\to b_3$}\\
  \textcolor{neqsstSThree}{$s_3\to b_2$}};
\draw[->,thick] (l1.east) -- (l2.west);
\draw[->,thick] (l2.east) -- (l3.west);
\node[hand] at (3.90,.85)
 {$s_1:\ b_1\to b_3$ costs $6$; \qquad
   $s_3:\ b_2\to b_2$ costs $1$};
\node[score,fill=orange!9] (ls) at (3.90,-.55)
 {Assignment reward $=30$; handover cost $=7$\\[2pt]
  $H_{\mathrm{local}}=30-7=23$};

\begin{scope}[xshift=9.45cm]
\node[font=\bfseries] at (3.90,3.83) {Globally coupled optimum};
\node[segbox] (g1) at (1.20,2.50)
 {$k=1$\\[3pt]\textcolor{neqsstSTwo}{$s_2\to b_1$}};
\node[segbox] (g2) at (3.90,2.50)
 {$k=2$\\[3pt]\textcolor{neqsstSTwo}{$s_2\to b_1$}\\
  \textcolor{neqsstSThree}{$s_3\to b_2$}};
\node[segbox] (g3) at (6.60,2.50)
 {$k=3$\\[3pt]\textcolor{neqsstSOne}{$s_1\to b_3$}\\
  \textcolor{neqsstSThree}{$s_3\to b_2$}};
\draw[->,thick] (g1.east) -- (g2.west);
\draw[->,thick] (g2.east) -- (g3.west);
\node[hand] at (3.90,.85)
 {$s_2:\ b_1\to b_1$ costs $1$; \qquad
   $s_3:\ b_2\to b_2$ costs $1$};
\node[score,fill=green!10] (gs) at (3.90,-.55)
 {Assignment reward $=28$; handover cost $=2$\\[2pt]
  $H^\star=28-2=26$};
\end{scope}

\draw[->,line width=1.3pt,green!45!black]
 (ls.east) -- node[above,font=\bfseries] {$+3$} (gs.west);
\node[font=\footnotesize,green!35!black] at (8.60,-1.36)
 {global objective gain};
\node[draw,rounded corners,fill=white,inner sep=4pt,
  font=\footnotesize] at (8.65,-2.03)
 {\textcolor{neqsstSOne}{$\blacksquare$} surfer $s_1$\qquad
  \textcolor{neqsstSTwo}{$\blacksquare$} surfer $s_2$\qquad
  \textcolor{neqsstSThree}{$\blacksquare$} surfer $s_3$};
\end{tikzpicture}%
}
\caption{Independent segmentwise optimization versus the globally coupled multi-segment optimum. The independent schedule achieves the larger raw assignment reward but incurs a handover cost of $7$. The globally coupled schedule sacrifices two reward units and lowers the handover cost to $2$, increasing the objective from $23$ to $26$.}
\label{fig:local-versus-global}
\end{figure}

\subsection{Depth and Resource Analysis}
\label{sec:resource-analysis}

The circuit diagram writes each CE-QAOA cost layer as the global phase unitary
\[
U_C(\gamma_\ell)=\exp[-i\gamma_\ell H_C].
\]
For the multi-segment WaaS instance, \(H_C\) is obtained from the objective in Eq.~\eqref{eq:waas-pauli-z-cost}. This compact notation represents a
phase operator over all segment--breaker blocks. Naively, the diagram suggests a deep global circuit, since the objective contains interactions within segments and across adjacent road segments. The diagonal structure of Eq.~\eqref{eq:waas-pauli-z-cost} gives a
favorable implementation. Every summand is diagonal in the computational basis.
All cost terms commute. The phase separator therefore factors into elementary
diagonal rotations,
\[
U_C(\gamma_\ell)
=
\prod_r \exp[-i\gamma_\ell h_r],
\qquad [h_r,h_{r'}]=0 ,
\]
where the factors \(h_r\) are the linear assignment terms, segmentwise
collision terms, and handover terms appearing in
Eq.~\eqref{eq:waas-pauli-z-cost}. The compiler can group, reorder,
route, and parallelize these rotations according to the hardware connectivity
and transpilation schedule.

In the dense case with \(K\) segments, \(B\) breakers per segment, and \(S\)
actual surfer choices per block (excluding the dummy label), the direct
cost-phase term counts in the binary formulation are
\[
N_{\mathrm{lin}} = KBS,\qquad
N_{\mathrm{surf}} = KS\binom{B}{2},\qquad
N_{\mathrm{break}} = KB\binom{S}{2},\qquad
N_{\mathrm{hand}} = |E_{\mathrm{seg}}|SB^2 .
\]

These counts give the elementary diagonal phase workload. The realized depth is
controlled by the interaction graph, hardware routing, edge coloring, and
parallel scheduling of the commuting phase rotations.

The CE-QAOA encoding reduces this workload by absorbing breaker capacity into
the Hilbert space. Each segment--breaker block is one-hot, so a breaker on a
given segment can select at most one surfer by construction. Hence the
\(P_2\)-family in Eq.~\eqref{eq:waas-pauli-z-cost}, with
\(KB\binom{S}{2}\) pairwise terms in the dense case, vanishes on the encoded
subspace and can be omitted from the implemented phase Hamiltonian. The cost
phase then focuses on assignment rewards, surfer uniqueness across breaker
blocks, and handover coupling across adjacent segments.

The initialization depth has a separate and striking scaling. Each block is
prepared as a one-hot \(W\)-state. A single dense $(S+1)$-qubit block can be prepared
ancilla-free using $S$ suitable excitation-preserving two-qubit rotations on a line,
and all segment--breaker blocks can be prepared in parallel \cite{onahce}.
Thus the full initialization depth is the depth of one block preparation,
rather than the depth of a serial preparation over all \(KB\) blocks. The mixer has the same block-depth property and all blocks can run in parallel. The mixer depth is therefore controlled by the implementation depth of one block-local XY mixer,
with efficient realizations using native \(\mathrm{iSWAP}^{\alpha}\),
\(f\mathrm{Sim}\), \(XX+YY\)-type interactions, or standard compiled
two-qubit primitives \cite{onahce}.

This gives the resource picture for the multi-segment CE-QAOA circuit. The
global phase layer is a commuting diagonal phase network with flexible
scheduling. The state preparation depth and mixer depth are limited by the
single-block implementation. The encoded one-hot structure removes an entire
breaker-capacity penalty family from the direct Farhi-type QAOA phase
Hamiltonian, reducing the diagonal phase workload while preserving the
assignment structure needed for the WaaS objective \cite{onahce}. Consequently, the
multi-segment CE-QAOA circuit combines a commuting diagonal cost phase with
parallel block-local state preparation and mixing desirable for near-term devices because it combines shallowness with robust performance guarantees\cite{onahce, onahfinite}. The one-hot encoding also separates transport from phase selection. The
block-local XY mixer controls probability transport inside each
single-excitation subspace, while the diagonal cost unitary assigns phases to
the decoded WaaS schedules. The precise primitivity conditions for each one-hot block follow below.
The cost angles perform instance-dependent phase selection over assignment
rewards, surfer-collision penalties, and handover terms.

\begin{proposition}[Primitivity of a complete-XY one-hot block]
\label{prop:waas-mixer-primitivity}
Let $d\geq2$ and use the physical mixer normalization in
\eqref{eq:waas-mixer-singleton}. On the one-excitation sector,
$\widetilde H_{XY}=2(J_d-I_d)/(d-1)$. The stochasticized
transition probabilities satisfy
\begin{align}
P_{uv}(\beta)
&=\frac{4}{d^2}\sin^2\left(\frac{d\beta}{d-1}\right)
&& (u\ne v),\label{eq:waas-mixer-offdiag}\\
P_{uu}(\beta)
&=1-\frac{4(d-1)}{d^2}
\sin^2\left(\frac{d\beta}{d-1}\right).
\label{eq:waas-mixer-diag}
\end{align}
For $d\geq3$, the kernel is primitive if and only if
$d\beta/(d-1)\notin\pi\mathbb Z$. For $d=2$, it is primitive
if and only if $\beta\notin(\pi/4)\mathbb Z$.
Consequently $\beta=\pi/8$ makes every nontrivial block kernel
strictly positive, regardless of block size.
\end{proposition}
\begin{proof}
On the one-hot sector, $X_aX_b+Y_aY_b$ has off-diagonal
matrix element $2$ between $|e_a\rangle$ and $|e_b\rangle$.
Thus $\widetilde H_{XY}=2(J_d-I_d)/(d-1)$ with eigenvalues
$2$ and $-2/(d-1)$. Spectral decomposition gives the displayed
transition probabilities. For $d\geq3$ the diagonal probability
is at least $(d-2)^2/d^2>0$; the off-diagonal one is positive
exactly under the stated condition. Hence the matrix is positive
and primitive. For $d=2$ the transition matrix is
\[
\begin{pmatrix}
\cos^2(2\beta)&\sin^2(2\beta)\\
\sin^2(2\beta)&\cos^2(2\beta)
\end{pmatrix},
\]
which is primitive precisely when both entries are positive.
The choice $\beta=\pi/8$ satisfies these conditions for every $d\geq2$.
\end{proof}

\subsection{Implementation}
\label{sec:implementation}

We implemented the multi-segment WaaS formulation as a hybrid
classical--quantum validation pipeline. The implementation consists of four
components: synthetic instance generation, a direct classical reference solver,
an exact nested \textsc{NEQSST} solver, and a Qiskit implementation of the
nested quantum sampling circuit. The purpose of this pipeline is to verify that
the nested one-hot formulation represents the same optimization problem as the
direct multi-segment formulation, and to evaluate the schedules produced by the
quantum sampler in the same objective space.

\paragraph{Quantum circuit depth.}
All numerical experiments reported in this work employ
a single-layer ($p=1$) NEQSST circuit,
\begin{equation}
\label{eq:neqsst_single_layer}
|\psi_1(\gamma,\beta)\rangle
=
U_M(\beta)U_C(\gamma)|s_0\rangle.
\end{equation}
Thus, each quantum sampling experiment uses one
cost-phase separator and one block-local XY mixer,
with two variational parameters $(\gamma,\beta)$.

The three-layer circuit shown in
Fig.~\ref{fig:neqsst_circuit} illustrates the
general depth-$p$ architecture and should not be
interpreted as the circuit depth used in the
numerical experiments.

Synthetic instances are generated with
\[
S\in\{2,3,4,5\}, \qquad B\in\{2,3,4,5\}, \qquad K\in\{1,2,3\},
\]
where \(S\) is the number of surfers, \(B\) is the number of breakers, and
\(K\) is the number of road segments. For each surfer--breaker--segment triple
\((s,b,k)\), the data generator constructs admissibility indicators, assignment
rewards \(a_{s,b,k}\), timing penalties, and adjacent-segment handover costs
\(\Delta T_{b,b',k,k'}\). Each instance therefore contains the full
multi-segment structure of interest including: local assignment rewards, surfer uniqueness constraints within each segment, breaker capacity constraints within each segment, and handover couplings between adjacent road segments.

As a classical reference, we solve the direct hard-constrained WaaS optimization
problem with Gurobi. The Gurobi model uses binary variables \(x_{s,b,k}\) and
imposes the segmentwise constraints
\[
\sum_b x_{s,b,k}\leq 1,
\qquad
\sum_s x_{s,b,k}\leq 1.
\]
The hard-constrained reference also fixes $x_{s,b,k}=0$ for every
timing-infeasible triple. Its objective contains assignment rewards and
adjacent-segment handover penalties. This produces the reference objective value
\(H^\star_{\mathrm{ref}}\) and the corresponding optimal WaaS schedule.

\section{Results}
\label{sec:results}

The nested \textsc{NEQSST} solver represents the same problem using
segment--breaker one-hot blocks. For each admissible segment--breaker pair
\((k,b)\), the block \(Y_{k,b}\) contains one dummy label for no drafting and
one label for each admissible surfer. Thus block label \(0\) represents no
surfer assigned to breaker \(b\) on segment \(k\), while the surfer label \(s\)
represents the assignment \((s,b,k)\). In the dense case, the nested register
contains
\[
N_{\mathrm{qubits}} = KB(S+1)
\]
qubits, where the additional state in each block represents no drafting. The
exact nested solver optimizes over these block labels and decodes them into the
classical variables \(x_{s,b,k}\). This gives an encoded objective value
\(H^\star_{\mathrm{enc}}\). The equality
\[
H^\star_{\mathrm{enc}} = H^\star_{\mathrm{ref}}
\]
is proved at the formulation level under the explicit penalty conditions
of Propositions~\ref{prop:waas-penalty-exactness} and
\ref{prop:waas-encoding-equivalence}. The reported solver agreement
validates the numerical implementation of that exact equivalence.

For debugging and baseline comparison, we include a nested random sampler. This
sampler draws one label from each segment--breaker block, decodes the sampled
labels into \(x_{s,b,k}\), repairs any surfer collision by keeping the locally
best assignment for each pair \((s,k)\), and evaluates the resulting schedule
under the WaaS objective. This sampler is a classical nested baseline used to
test the decoder and objective-evaluation pipeline.

The Qiskit implementation constructs the \textsc{NEQSST} circuit using the same
nested block structure. Each block is initialized in a one-hot \(W\)-state. The
mixer is a product of block-local complete-graph XY mixers, so the one-hot
constraint is preserved inside every segment--breaker block. The diagonal cost
phase is built from the multi-segment objective and includes assignment rewards,
surfer uniqueness penalties, and adjacent-segment handover couplings. After
measurement, sampled bitstrings are decoded blockwise into classical WaaS
schedules and evaluated using the same objective function as the Gurobi and
exact encoded solvers.

\begin{table}[t]
\centering
\small
\setlength{\tabcolsep}{4pt}
\caption{
Validation of the nested \textsc{NEQSST} implementation on small multi-segment WaaS instances. The table compares the direct Gurobi reference
solver with the exact nested one-hot \textsc{NEQSST} formulation and the Qiskit implementation of the nested quantum sampler. The equality column verifies that the encoded search space preserves the direct WaaS optimum, while the Qiskit
columns report the best decoded objective found from finite-shot sampling. 
All Qiskit sampling results were obtained using
single-layer $p=1$ NEQSST circuits.
}
\label{tab:neqsst-validation}
\resizebox{\linewidth}{!}{%
\begin{tabular}{cccccccccccc}
\toprule
\(S\) & \(B\) & \(K\) & \(N_q\) &
Ref. solver & \(H^\star_{\mathrm{ref}}\) &
Encoded solver & \(H^\star_{\mathrm{enc}}\) &
Match &
Qiskit sampler & \(H_{\mathrm{q}}\) & Gap \\
\midrule
3 & 3 & 2 & 24 & Gurobi & 11.873 & GurobiNEQSST & 11.873 & True & NEQSST  & 11.873 & 0.000 \\
3 & 2 & 3 & 24 & Gurobi & 5.348 & GurobiNEQSST & 5.348 & True & NEQSST  & 5.348 & 0.000 \\
4 & 2 & 2 & 20 & Gurobi & 6.016 & GurobiNEQSST & 6.016 & True & NEQSST  & 6.016 & 0.000 \\
4 & 3 & 2 & 30 & Gurobi & 4.623 & GurobiNEQSST & 4.623 & True & NEQSST  & 4.623 & 0.000 \\
5 & 3 & 1 & 18 & Gurobi & 4.546 & GurobiNEQSST & 4.546 & True & NEQSST  & 4.546 & 0.000 \\
\bottomrule
\end{tabular}%
}
\end{table}

Table~\ref{tab:neqsst-validation} summarizes the implementation
validation. For all tested instances, the exact nested \textsc{NEQSST}
formulation attains the same objective value as the direct Gurobi reference
solver. This confirms that the segment--breaker one-hot encoding preserves the
hard-constrained WaaS optimum. The Qiskit sampler also recovers the optimum in
these small finite-shot tests, giving zero observed sampling gap on the reported
instances.

\section{Discussion and Limitations}
\label{sec:discussion}

The nested representation separates local capacity from global coupling.
Breaker capacity is encoded inside each segment--breaker block, while surfer
uniqueness and handover consistency remain in the phase Hamiltonian. This
separation removes one penalty family and allows block-local preparation and
mixing to proceed in parallel.

Route preprocessing is an important part of the construction. Vehicles that do
not traverse a segment generate neither assignment variables nor quantum
labels. The reduction can be substantial when vehicles enter and leave at
different junctions. It also prevents undefined segment-arrival times from
appearing in the objective.

The handover model captures continuity through the coefficient
\(1+|T_{b,k}-T_{b',k'}|\). The present convention charges a baseline cost for
every adjacent-segment continuation and adds timing mismatch when the selected
breaker changes. Other applications may prefer a switching-only coefficient.
Such alternatives change the objective coefficients without changing the
nested encoding.

The finite-shot decoder uses a local repair rule when one surfer is selected by
several breaker blocks on the same segment. This rule is computationally light,
but it does not account for how the retained breaker affects neighboring
handover terms. A handover-aware repair or a small classical postprocessing optimization can improve decoded schedules. The exact nested Gurobi validation does not use this heuristic and therefore remains the appropriate test of formulation equivalence. The complete-graph XY mixer requires interactions between every pair of labels inside a block. Native all-to-all connectivity is uncommon, so hardware realization may require swap networks or alternative sparse mixers. The commuting phase terms also require routing when the hardware graph does not match the problem interaction graph.

The reported simulations validate the nested encoding, cost construction, and decoding pipeline on small multi-segment instances; extending the implementation to larger route networks and quantum hardware remains a direction for future work. The present model also assumes deterministic travel times, fixed vehicle roles, and known segment availability. Extensions can include uncertain travel times, probabilistic route choice, capacity greater than one, role changes between segments, and general segment graphs with merges and branches.

\subsection{Conclusion}
\label{sec:conclusion}

We formulated multi-segment shared transportation as a coupled assignment problem with route availability, timing compatibility, segmentwise matching, and adjacent-segment handover costs. The formulation extends the single-segment maximum-weight matching model to routes on which vehicles enter and leave at different junctions. The \textsc{NEQSST} construction represents every admissible segment--breaker pair by a one-hot block containing the no-drafting state and
the available surfer labels. This nested encoding is in bijection with the breaker-capacity-feasible binary assignments; the associated penalty vanishes identically on the encoded Hilbert space and can therefore be removed from the implemented phase Hamiltonian without altering the constrained problem. Surfer uniqueness and handover structure remain as commuting diagonal couplings between blocks, while state preparation and XY mixing are parallel across blocks.

Under the stated sufficient penalty bounds, the nested and original hard-constrained formulations have exactly the same global optimum for every admissible instance. Exact nested optimization also matches the direct Gurobi formulation on all tested instances. The finite-shot Qiskit implementation recovers the same optima on
the reported small problems and produces decoded distributions over classical multi-segment schedules. Resource and sensitivity studies further show how
route preprocessing reduces the register and how the handover weight changes the globally optimal assignment pattern.

These results establish \textsc{NEQSST} as a hybrid quantum-classical sampling architecture for shared transportation problems. Future work will study larger problem instances, noisy quantum hardware, and segment graphs containing merges and branches.

\appendix


\appendix
\section{Analytically Solvable Multi-Segment WaaS Instances}
\label{sec:worked-examples}

We construct two small instances for which every stage of the
multi-segment formulation can be evaluated explicitly.

The first example contains two road segments and isolates the
interaction between route availability, segmentwise matching,
timing penalties, handover costs, and the nested one-hot
encoding.

The second example contains three road segments separated by
four junctions. Vehicles enter and leave the corridor at
different junctions, and the globally optimal schedule differs
from the collection of independently optimal segmentwise
matchings.

Together, these examples provide exact checks of the
multi-segment objective, the QUBO representation, the
constraint-preserving encoding, and the blockwise decoder.

\subsection{Example 1: Two Segments with One Intermediate
Intake and One Drop-Off}
\label{subsec:worked-example-two-segments}

Consider the linear corridor
\[
v_0
\xrightarrow{\;k=1\;}
v_1
\xrightarrow{\;k=2\;}
v_2.
\]

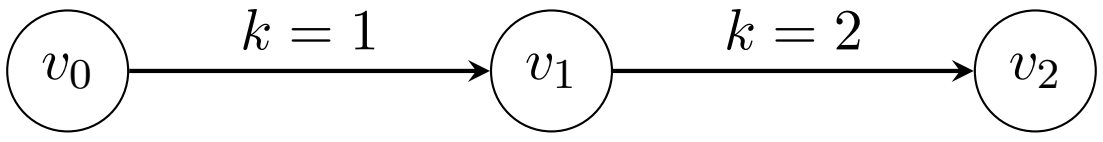
\begin{figure}[H]
\centering
\begin{tikzpicture}[>=stealth]

\node[circle,draw,minimum size=8mm]
    (v0) at (0,0) {$v_0$};

\node[circle,draw,minimum size=8mm]
    (v1) at (3.2,0) {$v_1$};

\node[circle,draw,minimum size=8mm]
    (v2) at (6.4,0) {$v_2$};

\draw[->,thick]
    (v0) -- node[above] {$k=1$} (v1);

\draw[->,thick]
    (v1) -- node[above] {$k=2$} (v2);

\node[anchor=west] at (0,-0.9)
    {$s_1:\ v_0\rightarrow v_1\rightarrow v_2$};

\node[anchor=west] at (0,-1.4)
    {$s_2:\ v_1\rightarrow v_2$};

\node[anchor=west] at (0,-1.9)
    {$b_1:\ v_0\rightarrow v_1\rightarrow v_2$};

\node[anchor=west] at (0,-2.4)
    {$b_2:\ v_0\rightarrow v_1$};

\end{tikzpicture}

\caption{
Two-segment corridor for Example~1.
Surfer $s_2$ enters at the intermediate junction
$v_1$, while breaker $b_2$ leaves the corridor at $v_1$.
}
\label{fig:example-one-routes}
\end{figure}

The route structure is shown in
Fig.~\ref{fig:example-one-routes}.

Surfer $s_1$ and breaker $b_1$ traverse the complete
corridor. Breaker $b_2$ leaves at $v_1$, while surfer
$s_2$ enters at $v_1$.

\subsubsection{Availability preprocessing}

The surfer- and breaker-availability matrices are
\[
A^S=
\begin{array}{c|cc}
 & k=1 & k=2\\
\hline
s_1 & 1 & 1\\
s_2 & 0 & 1
\end{array},
\qquad
A^B=
\begin{array}{c|cc}
 & k=1 & k=2\\
\hline
b_1 & 1 & 1\\
b_2 & 1 & 0
\end{array}.
\]

Classical route preprocessing retains only the four
assignment variables
\[
x_{s_1,b_1,1},
\qquad
x_{s_1,b_2,1},
\qquad
x_{s_1,b_1,2},
\qquad
x_{s_2,b_1,2}.
\]

All other surfer--breaker--segment triples are
unavailable and are removed.

\paragraph{Assignment rewards and timing penalties}

Let $\lambda_5=2$. Choose the assignment rewards and timing indicators
shown in Table~\ref{tab:example-one-rewards}.

\begin{table}[H]
\centering
\caption{
Assignment rewards and timing penalties for Example~1.
The effective reward is
$a_{s,b,k}=\widetilde w_{s,b,k}
-\lambda_5\delta_{s,b,k}$.
}
\label{tab:example-one-rewards}

\begin{tabular}{ccccc}
\toprule
Segment & Assignment
& $\widetilde w_{s,b,k}$
& $\delta_{s,b,k}$
& $a_{s,b,k}$\\
\midrule
$1$ & $(s_1,b_1)$ & $5$ & $0$ & $5$\\
$1$ & $(s_1,b_2)$ & $6$ & $0$ & $6$\\
$2$ & $(s_1,b_1)$ & $6$ & $0$ & $6$\\
$2$ & $(s_2,b_1)$ & $5$ & $1$ & $3$\\
\bottomrule
\end{tabular}
\end{table}

The effective assignment reward is
\begin{equation}
a_{s,b,k}
=
\widetilde w_{s,b,k}
-
\lambda_5\delta_{s,b,k}.
\label{eq:example-one-effective-reward}
\end{equation}

The final row illustrates the timing penalty:
the raw reward $5$ is reduced to the effective
reward $3$.

\paragraph{Timing-feasibility convention.}

The value $\lambda_5=2$ is deliberately chosen
to illustrate a \emph{soft} timing penalty. Accordingly, the objective used in this example
permits a timing-incompatible assignment but
reduces its reward. This illustrative penalty does not satisfy the
general sufficient condition for enforcing timing
feasibility as a hard constraint. The segmentwise matchings enumerated below are
feasible with respect to surfer uniqueness and
breaker capacity. Three of these schedules contain the timing-incompatible
assignment $(s_2,b_1,2)$. If timing compatibility is imposed as a hard constraint,
these three schedules must additionally be excluded. The optimum is unchanged under either convention,
as verified below. When hard timing feasibility is implemented by
removing incompatible labels during preprocessing,
the corresponding assignment variable does not
appear in the quantum register.

\subsubsection{Handover coefficients}

Set
\(
\lambda_h=1.
\)
For transparent arithmetic, choose the normalized
breaker timing marks
\[
T_{b_1,1}=0,
\qquad
T_{b_1,2}=0,
\qquad
T_{b_2,1}=3.
\]

The two possible continuation coefficients for
surfer $s_1$ are
\begin{align}
\Delta T_{b_1,b_1,1,2}
&=
1+\lvert 0-0\rvert
=1,
\\
\Delta T_{b_2,b_1,1,2}
&=
1+\lvert 3-0\rvert
=4.
\end{align}

Continuation behind $b_1$ therefore incurs
the baseline cost $1$, whereas changing from
$b_2$ to $b_1$ incurs cost $4$.

These coefficients follow the handover convention
of the main formulation, in which even a
same-breaker continuation incurs the baseline cost.

\paragraph{Complete binary objective and QUBO}

Introduce the compact notation
\[
x_1=x_{s_1,b_1,1},
\qquad
x_2=x_{s_1,b_2,1},
\]
\[
x_3=x_{s_1,b_1,2},
\qquad
x_4=x_{s_2,b_1,2}.
\]

Choose
\[
\lambda_3=\lambda_4=7,
\]
which exceeds the largest effective linear reward.

The complete maximization objective is
\begin{equation}
\label{eq:example-one-objective}
\boxed{
\begin{aligned}
H(\mathbf x)
={}&
5x_1+6x_2+6x_3+3x_4
\\
&-7x_1x_2
-7x_3x_4
\\
&-x_1x_3
-4x_2x_3.
\end{aligned}
}
\end{equation}

The term
\[
-7x_1x_2
\]
penalizes a surfer collision on segment $1$,
since $s_1$ cannot draft behind both breakers.

The term
\[
-7x_3x_4
\]
penalizes a breaker-capacity violation on
segment $2$, since breaker $b_1$ cannot
serve both surfers simultaneously.

The last two terms represent the continuation
costs incurred by surfer $s_1$.

Writing
\begin{equation}
H(\mathbf x)
=
\mathbf x^\top Q_{\max}\mathbf x,
\end{equation}
the symmetric maximization QUBO matrix is
\begin{equation}
\label{eq:example-one-qubo}
\boxed{
Q_{\max}
=
\begin{pmatrix}
5 & -\frac72 & -\frac12 & 0\\[2pt]
-\frac72 & 6 & -2 & 0\\[2pt]
-\frac12 & -2 & 6 & -\frac72\\[2pt]
0 & 0 & -\frac72 & 3
\end{pmatrix}.
}
\end{equation}

Indeed,
\[
\mathbf x^\top Q_{\max}\mathbf x
=
\sum_i(Q_{\max})_{ii}x_i
+
2\sum_{i<j}(Q_{\max})_{ij}x_ix_j,
\]
which reproduces
Eq.~\eqref{eq:example-one-objective} exactly.

For a minimization QUBO, use
\begin{equation}
Q_{\min}=-Q_{\max}.
\end{equation}

\subsubsection{Enumeration and global optimum}

The segmentwise matching choices on segment $1$ are
\[
\varnothing,
\qquad
(s_1,b_1),
\qquad
(s_1,b_2).
\]

The segmentwise matching choices on segment $2$ are
\[
\varnothing,
\qquad
(s_1,b_1),
\qquad
(s_2,b_1).
\]

Their Cartesian product therefore contains
\[
3\times3=9
\]
schedules satisfying both segmentwise matching
constraints. The complete enumeration is given in
Table~\ref{tab:example-one-enumeration}.

\begin{table}[H]
\centering
\caption{
Complete enumeration of the nine schedules satisfying
segmentwise surfer uniqueness and breaker capacity
in Example~1.
The handover cost is present only when surfer $s_1$
is assigned on both segments.
Schedules containing $(s_2,b_1)$ on segment $2$
are timing-incompatible under the hard-timing convention.
}
\label{tab:example-one-enumeration}

\begin{tabular}{ccc}
\toprule
Segment $1$ & Segment $2$ & $H$\\
\midrule

$\varnothing$
& $\varnothing$
& $0$\\

$\varnothing$
& $(s_1,b_1)$
& $6$\\

$\varnothing$
& $(s_2,b_1)$
& $3$\\

$(s_1,b_1)$
& $\varnothing$
& $5$\\

$(s_1,b_1)$
& $(s_1,b_1)$
& $5+6-1=10$\\

$(s_1,b_1)$
& $(s_2,b_1)$
& $5+3=8$\\

$(s_1,b_2)$
& $\varnothing$
& $6$\\

$(s_1,b_2)$
& $(s_1,b_1)$
& $6+6-4=8$\\

$(s_1,b_2)$
& $(s_2,b_1)$
& $6+3=9$\\

\bottomrule
\end{tabular}
\end{table}

The unique global maximizer is
\begin{equation}
\boxed{
\mathcal M^\star
=
\left\{
(s_1,b_1,1),
(s_1,b_1,2)
\right\},
\qquad
H^\star=10.
}
\label{eq:example-one-optimum}
\end{equation}

The locally larger reward
\[
a_{s_1,b_2,1}=6
\]
does not lead to the global optimum because
switching from $b_2$ to $b_1$ incurs the larger
handover cost $4$.

Under hard timing feasibility, the three schedules
containing $(s_2,b_1,2)$ are excluded.

The six remaining schedules have objective values
\[
0,\quad6,\quad5,\quad10,\quad6,\quad8.
\]

Thus the same unique optimum \eqref{eq:example-one-optimum} is obtained. This verifies directly that the formulation distinguishes local assignment rewards from global route-level optimization.

\subsubsection{Nested one-hot encoding}

Under the soft-timing convention, the nested
encoding contains the three blocks
\[
Y_{1,b_1}=\{0,s_1\},
\]
\[
Y_{1,b_2}=\{0,s_1\},
\]
and
\[
Y_{2,b_1}=\{0,s_1,s_2\}.
\]

Hence the total physical register size is
\begin{equation}
\boxed{
N_{\mathrm{qubits}}
=
2+2+3=7.
}
\label{eq:example-one-qubits}
\end{equation}

The optimal encoded label tuple is
\[
\bigl(
Y_{1,b_1},
Y_{1,b_2},
Y_{2,b_1}
\bigr)
=
(s_1,0,s_1).
\]

Its corresponding one-hot computational state is
\begin{equation}
\lvert e_{s_1}\rangle_{1,b_1}
\otimes
\lvert e_0\rangle_{1,b_2}
\otimes
\lvert e_{s_1}\rangle_{2,b_1}.
\label{eq:example-one-encoded-optimum}
\end{equation}

Breaker capacity on segment $2$ is enforced
inside the block $Y_{2,b_1}$.

Therefore,
\[
\boxed{
x_3x_4=0
}
\]
on every valid encoded basis state.

The penalty term
\[
-\lambda_4x_3x_4
\]
thus vanishes identically on the encoded subspace
and does not have to be implemented in the
quantum phase separator.

By contrast, surfer uniqueness on segment $1$
couples two different blocks,
\[
Y_{1,b_1}
\quad\text{and}\quad
Y_{1,b_2}.
\]

The product
\[
x_1x_2
\]
can be nonzero on an encoded basis state, so the
corresponding $\lambda_3$ penalty remains in the
encoded Hamiltonian.

This illustrates exactly which constraint
family is eliminated by the one-hot encoding
and which global relations must remain in
the phase Hamiltonian.

If hard timing compatibility is enforced by
excluding the label $s_2$ from $Y_{2,b_1}$,
that block contains only
\[
\{0,s_1\},
\]
and the register size becomes
\[
N_{\mathrm{qubits}}^{\mathrm{hard}}=2+2+2=6.
\]

The optimal encoded schedule is unchanged.

\subsection{Example 2: Three Segments with Staggered
Intake and Drop-Off}
\label{subsec:worked-example-three-segments}

Consider the corridor
\[
v_0
\xrightarrow{\;k=1\;}
v_1
\xrightarrow{\;k=2\;}
v_2
\xrightarrow{\;k=3\;}
v_3.
\]

Four junctions and three road segments are sufficient
to represent intermediate intake and intermediate
drop-off at distinct locations.

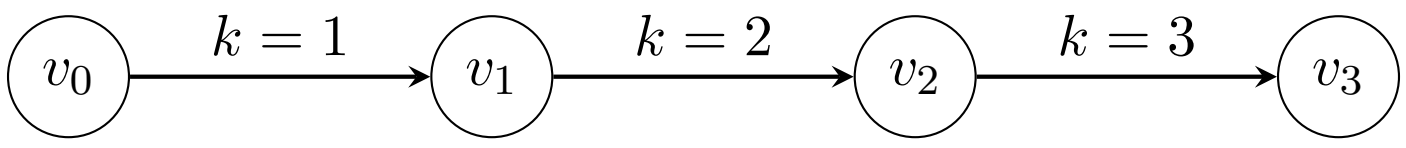
\begin{figure}[H]
\centering

\begin{tikzpicture}[>=stealth]

\node[circle,draw,minimum size=8mm]
    (v0) at (0,0) {$v_0$};

\node[circle,draw,minimum size=8mm]
    (v1) at (2.8,0) {$v_1$};

\node[circle,draw,minimum size=8mm]
    (v2) at (5.6,0) {$v_2$};

\node[circle,draw,minimum size=8mm]
    (v3) at (8.4,0) {$v_3$};

\draw[->,thick]
    (v0) -- node[above] {$k=1$} (v1);

\draw[->,thick]
    (v1) -- node[above] {$k=2$} (v2);

\draw[->,thick]
    (v2) -- node[above] {$k=3$} (v3);

\node[anchor=west] at (0,-0.9)
    {$s_1:\ v_0\rightarrow v_1\rightarrow v_2
             \rightarrow v_3$};

\node[anchor=west] at (0,-1.4)
    {$s_2:\ v_0\rightarrow v_1\rightarrow v_2$};

\node[anchor=west] at (0,-1.9)
    {$s_3:\ v_1\rightarrow v_2\rightarrow v_3$};

\node[anchor=west] at (0,-2.4)
    {$b_1:\ v_0\rightarrow v_1\rightarrow v_2$};

\node[anchor=west] at (0,-2.9)
    {$b_2:\ v_1\rightarrow v_2\rightarrow v_3$};

\node[anchor=west] at (0,-3.4)
    {$b_3:\ v_2\rightarrow v_3$};

\end{tikzpicture}

\caption{
Three-segment corridor for Example~2.
Surfer $s_3$ and breaker $b_2$ enter at $v_1$.
Surfer $s_2$ and breaker $b_1$ leave at $v_2$,
while breaker $b_3$ enters at $v_2$.
}
\label{fig:example-two-routes}
\end{figure}

\subsubsection{Availability matrices}

The surfer-availability matrix is
\[
A^S=
\begin{array}{c|ccc}
 & k=1 & k=2 & k=3\\
\hline
s_1 & 1 & 1 & 1\\
s_2 & 1 & 1 & 0\\
s_3 & 0 & 1 & 1
\end{array}.
\]

The breaker-availability matrix is
\[
A^B=
\begin{array}{c|ccc}
 & k=1 & k=2 & k=3\\
\hline
b_1 & 1 & 1 & 0\\
b_2 & 0 & 1 & 1\\
b_3 & 0 & 0 & 1
\end{array}.
\]

Classical preprocessing removes every assignment
triple for which the surfer or breaker does not
traverse the corresponding segment.

\subsubsection{Effective reward matrices}

The effective reward matrices are
\[
A^{(1)}
=
\begin{array}{c|c}
 & b_1\\
\hline
s_1 & 3\\
s_2 & 5
\end{array},
\]
\[
A^{(2)}
=
\begin{array}{c|cc}
 & b_1 & b_2\\
\hline
s_1 & 6 & 5\\
s_2 & 4 & 3\\
s_3 & 2 & 6
\end{array},
\]
and
\[
A^{(3)}
=
\begin{array}{c|cc}
 & b_2 & b_3\\
\hline
s_1 & 5 & 7\\
s_3 & 6 & 4
\end{array}.
\]

Here,
\[
A^{(k)}_{s,b}=a_{s,b,k}.
\]

Unavailable rows and columns have already been
removed by preprocessing.

For this example, assume every retained assignment
is timing-compatible. Thus the displayed effective
rewards do not require any additional hard-timing
exclusions.

Set
\[
\lambda_h=1,
\qquad
\lambda_3=\lambda_4=8.
\]

Both collision penalties exceed the largest
effective assignment reward, which is $7$.

\subsubsection{Handover matrices}

Choose the normalized breaker timing marks
\[
T_{b_1,1}=0,
\qquad
T_{b_1,2}=0,
\]
\[
T_{b_2,2}=3,
\qquad
T_{b_2,3}=3,
\qquad
T_{b_3,3}=5.
\]

The handover matrix from segment $1$ to
segment $2$ is
\[
\Delta T^{(1,2)}
=
\begin{array}{c|cc}
 & b_1 & b_2\\
\hline
b_1 & 1 & 4
\end{array}.
\]

The handover matrix from segment $2$ to
segment $3$ is
\[
\Delta T^{(2,3)}
=
\begin{array}{c|cc}
 & b_2 & b_3\\
\hline
b_1 & 4 & 6\\
b_2 & 1 & 3
\end{array}.
\]

For example,
\[
\Delta T_{b_1,b_3,2,3}
=
1+|0-5|
=
6,
\]
while
\[
\Delta T_{b_2,b_2,2,3}
=
1+|3-3|
=
1.
\]

Hence, continuing behind the same breaker can
be substantially less expensive than switching
breakers between consecutive segments.

\subsubsection{Independent segmentwise optimization}

First consider optimizing each segment separately,
without accounting for handover costs.

The maximum-weight matching on segment $1$ is
\[
\mathcal M_1^{\mathrm{loc}}
=
\{(s_2,b_1)\},
\qquad
R_1=5.
\]

On segment $2$, the maximum-weight matching is
\[
\mathcal M_2^{\mathrm{loc}}
=
\{
(s_1,b_1),
(s_3,b_2)
\},
\]
with
\[
R_2=6+6=12.
\]

On segment $3$, the maximum-weight matching is
\[
\mathcal M_3^{\mathrm{loc}}
=
\{
(s_1,b_3),
(s_3,b_2)
\},
\]
with
\[
R_3=7+6=13.
\]

The total segmentwise reward is therefore
\[
R_{\mathrm{loc}}
=
5+12+13
=
30.
\]

However, surfer $s_1$ changes from breaker $b_1$
on segment $2$ to breaker $b_3$ on segment $3$,
incurring
\[
\Delta T_{b_1,b_3,2,3}=6.
\]

Surfer $s_3$ remains with breaker $b_2$ and incurs
\[
\Delta T_{b_2,b_2,2,3}=1.
\]

There is no handover charge for surfer $s_2$
between segments $1$ and $2$, because $s_2$
is not assigned on segment $2$ in the locally
optimized schedule.

The full multi-segment value is consequently
\begin{equation}
\boxed{
H_{\mathrm{loc}}
=
5+12+13-6-1
=
23.
}
\label{eq:example-two-local-value}
\end{equation}

Although each segment is individually optimal,
their combination need not maximize the
coupled route-level objective.

\subsubsection{An exact dynamic-programming reduction}

The globally optimal solution can be obtained by
conditioning on the matching $\mathcal M_2$
selected on the middle segment. Let
\[
R_k(\mathcal M_k)
=
\sum_{(s,b)\in\mathcal M_k}a_{s,b,k}
\]
denote the assignment reward on segment $k$.
For adjacent segments, define
\[
C_{k,k+1}(\mathcal M_k,\mathcal M_{k+1})
=
\sum_{\substack{
(s,b)\in\mathcal M_k\\
(s,b')\in\mathcal M_{k+1}
}}
\Delta T_{b,b',k,k+1}.
\]

Because the matching constraints are local to
each segment, the left and right optimization
problems become independent once the middle
matching is fixed.

Define the optimal left contribution
\begin{equation}
L(\mathcal M_2)
=
\max_{\mathcal M_1}
\left[
R_1(\mathcal M_1)
-
C_{12}(\mathcal M_1,\mathcal M_2)
\right],
\label{eq:example-two-left-dp}
\end{equation}
and the optimal right contribution
\begin{equation}
R(\mathcal M_2)
=
\max_{\mathcal M_3}
\left[
R_3(\mathcal M_3)
-
C_{23}(\mathcal M_2,\mathcal M_3)
\right].
\label{eq:example-two-right-dp}
\end{equation}

For each fixed middle matching, the optimal
global objective is
\begin{equation}
F(\mathcal M_2)
=
L(\mathcal M_2)
+
R_2(\mathcal M_2)
+
R(\mathcal M_2).
\label{eq:example-two-middle-value}
\end{equation}

Hence the global optimum is
\begin{equation}
H^\star
=
\max_{\mathcal M_2}
\left[
L(\mathcal M_2)
+
R_2(\mathcal M_2)
+
R(\mathcal M_2)
\right].
\label{eq:example-two-exact-dp}
\end{equation}

The middle segment has three surfers and two
breakers.

Its feasible matchings consist of:

\begin{itemize}
\item one empty matching;
\item six one-assignment matchings;
\item six two-assignment matchings.
\end{itemize}

There are therefore exactly
\(
1+6+6=13
\)
middle-segment matchings.
The first segment has three feasible matchings,
and the third has seven.
Consequently, the complete feasible search
contains
\(
3\times13\times7
=
273
\)
combinations of segmentwise matchings.
The dynamic-programming decomposition evaluates
the same optimization by conditioning on only
thirteen middle-segment states.

\subsubsection{Complete middle-segment enumeration}

All thirteen possible middle-segment matchings
and their optimal left and right continuations
are listed in
Table~\ref{tab:example-two-middle-dp}.

\begin{table}[H]
\centering
\scriptsize

\caption{
The middle-segment dynamic-programming calculation
for Example~2.
$L$ is the optimal contribution from segment $1$,
including the $1\rightarrow2$ handover cost.
$R$ is the optimal contribution from segment $3$,
including the $2\rightarrow3$ handover cost.
}
\label{tab:example-two-middle-dp}

\resizebox{\linewidth}{!}{%
\begin{tabular}{ccccccc}
\toprule

$\mathcal M_2$
&
$R_2(\mathcal M_2)$
&
Best $\mathcal M_1$
&
$L(\mathcal M_2)$
&
Best $\mathcal M_3$
&
$R(\mathcal M_2)$
&
$F(\mathcal M_2)$
\\

\midrule

$\varnothing$
& $0$
& $\{s_2\!\to b_1\}$
& $5$
& $\{s_1\!\to b_3,s_3\!\to b_2\}$
& $13$
& $18$
\\

$\{s_1\!\to b_1\}$
& $6$
& $\{s_2\!\to b_1\}$
& $5$
& $\{s_1\!\to b_3,s_3\!\to b_2\}$
& $7$
& $18$
\\

$\{s_1\!\to b_2\}$
& $5$
& $\{s_2\!\to b_1\}$
& $5$
& $\{s_1\!\to b_3,s_3\!\to b_2\}$
& $10$
& $20$
\\

$\{s_2\!\to b_1\}$
& $4$
& $\{s_2\!\to b_1\}$
& $4$
& $\{s_1\!\to b_3,s_3\!\to b_2\}$
& $13$
& $21$
\\

$\{s_2\!\to b_2\}$
& $3$
& $\{s_1\!\to b_1\}$
& $3$
& $\{s_1\!\to b_3,s_3\!\to b_2\}$
& $13$
& $19$
\\

$\{s_3\!\to b_1\}$
& $2$
& $\{s_2\!\to b_1\}$
& $5$
& $\{s_1\!\to b_3,s_3\!\to b_2\}$
& $9$
& $16$
\\

$\{s_3\!\to b_2\}$
& $6$
& $\{s_2\!\to b_1\}$
& $5$
& $\{s_1\!\to b_3,s_3\!\to b_2\}$
& $12$
& $23$
\\

$\{s_1\!\to b_1,s_2\!\to b_2\}$
& $9$
& $\{s_1\!\to b_1\}$
& $2$
& $\{s_1\!\to b_3,s_3\!\to b_2\}$
& $7$
& $18$
\\

$\{s_1\!\to b_1,s_3\!\to b_2\}$
& $12$
& $\{s_2\!\to b_1\}$
& $5$
& $\{s_1\!\to b_3,s_3\!\to b_2\}$
& $6$
& $23$
\\

$\{s_1\!\to b_2,s_2\!\to b_1\}$
& $9$
& $\{s_2\!\to b_1\}$
& $4$
& $\{s_1\!\to b_3,s_3\!\to b_2\}$
& $10$
& $23$
\\

$\{s_1\!\to b_2,s_3\!\to b_1\}$
& $7$
& $\{s_2\!\to b_1\}$
& $5$
& $\{s_1\!\to b_3,s_3\!\to b_2\}$
& $6$
& $18$
\\

$\{s_2\!\to b_1,s_3\!\to b_2\}$
& $10$
& $\{s_2\!\to b_1\}$
& $4$
& $\{s_1\!\to b_3,s_3\!\to b_2\}$
& $12$
& $\mathbf{26}$
\\

$\{s_2\!\to b_2,s_3\!\to b_1\}$
& $5$
& $\{s_1\!\to b_1\}$
& $3$
& $\{s_1\!\to b_3,s_3\!\to b_2\}$
& $9$
& $17$
\\

\bottomrule
\end{tabular}%
}
\end{table}

The table evaluates the globally coupled
objective exactly.

In particular, the unique maximizer of
$F(\mathcal M_2)$ is
\[
\boxed{
\mathcal M_2^\star
=
\{
(s_2,b_1),
(s_3,b_2)
\},
}
\]
with
\[
F(\mathcal M_2^\star)=26.
\]

Its optimal left continuation is
\[
\mathcal M_1^\star
=
\{(s_2,b_1)\},
\]
and its optimal right continuation is
\[
\mathcal M_3^\star
=
\{
(s_1,b_3),
(s_3,b_2)
\}.
\]

\subsubsection{Globally optimal schedule}

Combining the three segmentwise schedules gives
the unique global optimum
\begin{equation}
\boxed{
\begin{aligned}
\mathcal M^\star
=
\{&
(s_2,b_1,1),
(s_2,b_1,2),
\\
&
(s_3,b_2,2),
(s_3,b_2,3),
\\
&
(s_1,b_3,3)
\}.
\end{aligned}
}
\label{eq:example-two-optimal-schedule}
\end{equation}

Its assignment reward is
\(
R^\star
=
5+(4+6)+(7+6)
=
28.
\)
There are two continuation costs:
\[
\Delta T_{b_1,b_1,1,2}=1
\]
for surfer $s_2$, and
\[
\Delta T_{b_2,b_2,2,3}=1
\]
for surfer $s_3$.
Thus
\begin{equation}
\boxed{
H^\star
=
5+(4+6)+(7+6)-1-1
=
26.
}
\label{eq:example-two-global-optimum}
\end{equation}

The global optimum retains surfer $s_2$
behind breaker $b_1$ from segment $1$ to
segment $2$. It also retains surfer $s_3$ behind breaker
$b_2$ from segment $2$ to segment $3$.
Surfer $s_1$ begins drafting only on
segment $3$. Relative to the independently optimized
schedule, the global solution sacrifices
two units of assignment reward on segment $2$:
\(
30-28=2.
\)
However, the total handover cost decreases from
$7$ to $2$, yielding a saving of
\(
7-2=5.
\)
The resulting net improvement is
\begin{equation}
H^\star-H_{\mathrm{loc}}
=
(28-2)-(30-7)
=
3.
\label{eq:example-two-global-improvement}
\end{equation}

This establishes explicitly that a globally
optimal multi-segment schedule need not be
obtained by independently optimizing each
segment.

\subsubsection{Nested quantum encoding}

After route preprocessing, the surviving
segment--breaker blocks are
\[
Y_{1,b_1}
=
\{0,s_1,s_2\},
\]
\[
Y_{2,b_1}
=
\{0,s_1,s_2,s_3\},
\]
\[
Y_{2,b_2}
=
\{0,s_1,s_2,s_3\},
\]
\[
Y_{3,b_2}
=
\{0,s_1,s_3\},
\]
and
\[
Y_{3,b_3}
=
\{0,s_1,s_3\}.
\]

The preprocessed quantum register size is
\begin{equation}
\boxed{
N_{\mathrm{qubits}}
=
3+4+4+3+3
=
17.
}
\label{eq:example-two-qubits}
\end{equation}

The corresponding dense register, before
route preprocessing, would contain
\(
KB(S+1)
=
3\cdot3\cdot4
=
36
\)
physical qubits. The reduction is therefore $36-17=19$ qubits.
This reduction follows solely from known
route-availability information, without
discarding any admissible assignment.

The optimal encoded label tuple is
\begin{equation}
\begin{aligned}
\bigl(
&Y_{1,b_1},
Y_{2,b_1},
Y_{2,b_2},
Y_{3,b_2},
Y_{3,b_3}
\bigr)
=
(s_2,s_2,s_3,s_3,s_1).
\end{aligned}
\label{eq:example-two-encoded-optimum}
\end{equation}
Its one-hot computational-basis representation is
\begin{equation}
\begin{aligned}
\lvert\mathcal M^\star\rangle
={}&
\lvert e_{s_2}\rangle_{1,b_1}
\otimes
\lvert e_{s_2}\rangle_{2,b_1}
\otimes
\lvert e_{s_3}\rangle_{2,b_2}
\\
&\otimes
\lvert e_{s_3}\rangle_{3,b_2}
\otimes
\lvert e_{s_1}\rangle_{3,b_3}.
\end{aligned}
\label{eq:example-two-optimal-basis-state}
\end{equation}

Blockwise decoding returns precisely the
schedule $\mathcal M^\star$ in
Eq.~\eqref{eq:example-two-optimal-schedule}.
Every breaker-capacity constraint is
satisfied by construction within the
corresponding one-hot block.
The remaining cross-block constraints are:

\begin{itemize}

\item surfer uniqueness among breaker blocks
on the same road segment;

\item continuation and handover interactions
between adjacent road segments.

\end{itemize}

These are precisely the relations retained
in the encoded diagonal phase Hamiltonian.

\subsection{Interpretation of the Worked Instances}
\label{subsec:worked-examples-interpretation}

The two examples isolate complementary
features of the NEQSST construction.

\paragraph{An exact encoding of local constraints.}

Example~1 demonstrates directly that breaker
capacity is enforced by the one-hot block
structure.
An assignment collision involving two surfers
and the same breaker cannot occur inside
a valid encoded block.
Its corresponding QUBO penalty therefore
vanishes identically on the encoded subspace.
In contrast, surfer uniqueness involves
different breaker blocks and must be
represented by an inter-block constraint
or eliminated by an additional classical
or quantum mechanism.

\paragraph{Nonseparability of the route objective.}

Example~2 demonstrates why the multi-segment
problem is not generally solved by independent
maximum-weight matchings.
The handover terms couple neighboring
segment decisions.
As a result, the optimal global schedule may
sacrifice local assignment reward to improve
continuation costs across the route.
The example exhibits the strict inequality
\[
H^\star=26>23=H_{\mathrm{loc}}.
\]

\paragraph{Preservation of the optimization problem.}

In both examples, the nested one-hot labels
represent the same assignment choices as the
corresponding breaker-capacity-constrained
binary formulation. The decoder maps each encoded label configuration
to an assignment schedule without introducing
breaker-capacity violations. Provided the remaining hard constraints are
properly enforced by preprocessing, sufficiently
large penalties, or exact feasibility handling,
the nested representation preserves the
optimum of the original multi-segment
optimization problem.

\paragraph{Role of the quantum sampler.}

The worked examples establish the classical
objective, the relevant configuration space,
and anoptimum independently of
a particular circuit parameterization. They therefore provide useful test cases
for verifying that the quantum phase Hamiltonian,
one-hot initialization, mixer, and decoder
represent the intended problem. They do not, by themselves, imply a
finite-depth probability guarantee for the
NEQSST quantum sampler.
Such a guarantee concerns the distribution
of measurement outcomes generated by the
chosen circuit and must be evaluated
separately from formulation equivalence.
Together, these examples connect the
multi-segment WaaS model to its nested
constraint-embedded representation and
provide transparent benchmarks for
the numerical results in the main text.

\printbibliography

@misc{onahce,
      title={Empirical Quantum Advantage in Constrained Optimization from Encoded Unitary Designs}, 
      author={Chinonso Onah and Roman Firt and Kristel Michielsen},
      year={2026},
      eprint={2511.14296},
      archivePrefix={arXiv},
      primaryClass={cs.ET},
      url={https://arxiv.org/abs/2511.14296}, 
}

@book{Burkard2009,
  author    = {Burkard, Rainer E. and Dell'Amico, Mauro and Martello, Silvano},
  title     = {Assignment Problems},
  publisher = {Society for Industrial and Applied Mathematics},
  address   = {Philadelphia, PA},
  year      = {2009},
  isbn      = {978-0-898716-63-4},
  doi       = {10.1137/1.9780898717762}
}

@article{Hadfield2019AOA,
  author  = {Stuart Hadfield and Zhihui Wang and Bryan O'Gorman and Eleanor G. Rieffel and Davide Venturelli and Rupak Biswas},
  title   = {From the Quantum Approximate Optimization Algorithm to a Quantum Alternating Operator Ansatz},
  journal = {Algorithms},
  volume  = {12},
  number  = {2},
  pages   = {34},
  year    = {2019},
  doi     = {10.3390/a12020034}
}

@article{alam2015review,
  author = {Alam, A. and Besselink, B. and Turri, V. and Mårtensson, J. and Johansson, K. H.},
  year = {2015},
  title = {Heavy-duty vehicle platooning for sustainable freight transportation: A review},
  journal = {IEEE Control Systems Magazine},
  volume = {35},
  number = {6},
  pages = {34-56},
}

@misc{onah2026solverzoo,
      title={Quantum and classical approaches to the optimization of highway platooning: the two-vehicle matching problem}, 
      author={Chinonso Onah and Agneev Guin and Carsten Othmer and J. A. Montañez-Barrera and Kristel Michielsen},
      year={2026},
      eprint={2603.18919},
      archivePrefix={arXiv},
      primaryClass={quant-ph},
      url={https://arxiv.org/abs/2603.18919}, 
}

@article{duoba2023empirical,
  title={Empirical equations of changes in aerodynamic drag based on direct on-track road load measurements for multi-vehicle platoons},
  author={Duoba, Michael},
  journal={SAE International Journal of Advances and Current Practices in Mobility},
  volume={5},
  number={2023-01-0830},
  pages={2344--2353},
  year={2023}
}

@article{Lucas2014,
  author  = {Lucas, Andrew},
  title   = {Ising Formulations of Many NP Problems},
  journal = {Frontiers in Physics},
  volume  = {2},
  pages   = {5},
  year    = {2014},
  doi     = {10.3389/fphy.2014.00005}
}

@article{Katzgraber2015,
  author  = {Katzgraber, Helmut G. and Andrist, Ruben S. and Hamze, Firas and Zhu, Zheng and Ochoa, Andrew J.},
  title   = {Avoiding Embedding Overhead in Quantum Annealing},
  journal = {Physical Review X},
  volume  = {5},
  number  = {3},
  pages   = {031032},
  year    = {2015},
  doi     = {10.1103/PhysRevX.5.031032}
}

@misc{onah2025quest,
  author        = {Onah, Chinonso and Miscasci, Neel and Othmer, Carsten and Michielsen, Kristel},
  title         = {{QUEST}: {QU}antum-{E}nhanced {S}hared {T}ransportation},
  year          = {2025},
  eprint        = {2505.08074},
  archivePrefix = {arXiv},
  primaryClass  = {quant-ph},
  url           = {https://arxiv.org/abs/2505.08074}
}

@misc{onahfinite,
      title={Finite-Depth, Finite-Shot Guarantees for Constrained Quantum Optimization via Fej\'er Filtering}, 
      author={Chinonso Onah and Kristel Michielsen},
      year={2026},
      eprint={2603.01809},
      archivePrefix={arXiv},
      primaryClass={quant-ph},
      url={https://arxiv.org/abs/2603.01809}, 
}

\end{document}